%% file: main.tex
\documentclass[10pt, letter, onecolumn]{arxiv}

\usepackage{kantlipsum, lipsum}
\usepackage{dm-colors}
\usepackage{amsmath}
\usepackage{pstricks, pst-node}
\usepackage{verbatim}
\usepackage{multirow}
\usepackage{scalerel}
\usepackage{booktabs}
\usepackage{enumitem}
\usepackage{xspace}
\usepackage{bm}
\usepackage{bbm}
\usepackage{mathtools}
\usepackage{soul}
\usepackage{epsfig}
\usepackage{graphicx}
\usepackage{tcolorbox}
\usepackage{subcaption}
\usepackage{amssymb}
\usepackage{colortbl}
\usepackage{csquotes}
\usepackage{etex}
\usepackage{setspace}
\usepackage{svg}
\usepackage{colortbl}
\usepackage{tabularx,ragged2e}
\usepackage{placeins}
\usepackage[symbol]{footmisc}
\usepackage[bibstyle=nature,citestyle=numeric-comp,%
            natbib=true,backend=biber,maxbibnames=99,%
            giveninits=false,sorting=none]{biblatex}
\usepackage{nameref}
\usepackage{varioref}
\usepackage[pagebackref=false,breaklinks=false,%
            colorlinks=true,bookmarks=true,citecolor=ourdarkblue,%
            urlcolor=ourdarkblue,linkcolor=ourdarkblue]{hyperref}
\usepackage[noabbrev,capitalize]{cleveref}
\usepackage{etoc}
\usepackage{longtable}

\graphicspath{{figures/}}

\title{{Exploring Large Language Models \\ for Specialist-level Oncology Care}}

\author[$\ast$,1]{Anil Palepu}
\author[$\ast$,3]{Vikram Dhillon}
\author[3]{Polly Niravath}
\author[1]{Wei-Hung Weng}
\author[3]{Preethi Prasad}
\author[2]{Khaled Saab}
\author[2]{Ryutaro Tanno}
\author[2]{Yong Cheng}
\author[3]{Hanh Mai}
\author[3]{Ethan Burns}
\author[3]{Zainub Ajmal}
\author[1]{Kavita Kulkarni}
\author[2]{\\Philip Mansfield}
\author[1]{Dale Webster}
\author[2]{Joelle Barral}
\author[1]{Juraj Gottweis}
\author[1]{Mike Schaekermann}
\author[2]{\\S. Sara Mahdavi}
\author[1]{Vivek Natarajan}
\author[1]{Alan Karthikesalingam}
\author[2]{Tao Tu}

\affil[1]{Google Research, }
\affil[2]{Google DeepMind, }
\affil[3]{Houston Methodist}

\renewcommand{\correspondingauthor}[1]{$\ast$~Equal contributions. %
                                       $\dagger$~Equal leadership. \\%
                                       $\ddagger$~Corresponding authors: dhillonv10@gmail.com, \{taotu, alankarthi, natviv\}@google.com }

\begin{document}

\begin{refsection}

\begin{abstract}
Large language models (LLMs) have shown remarkable progress in encoding clinical knowledge and responding to complex medical queries with appropriate clinical reasoning. However, their applicability in subspecialist or complex medical settings remains underexplored. In this work, we probe the performance of AMIE, a research conversational diagnostic AI system, in the subspecialist domain of breast oncology care without specific fine-tuning to this challenging domain. To perform this evaluation, we curated a set of 50 synthetic breast cancer vignettes representing a range of treatment-naive and treatment-refractory cases and mirroring the key information available to a multidisciplinary tumor board for decision-making (openly released with this work). We developed a detailed clinical rubric for evaluating management plans, including axes such as the quality of case summarization, safety of the proposed care plan, and recommendations for chemotherapy, radiotherapy, surgery and hormonal therapy. To improve performance, we enhanced AMIE with the inference-time ability to perform web search retrieval to gather relevant and up-to-date clinical knowledge and refine its responses with a multi-stage self-critique pipeline. We compare response quality of AMIE with internal medicine trainees, oncology fellows, and general oncology attendings under both automated and specialist clinician evaluations. In our evaluations, AMIE outperformed trainees and fellows demonstrating the potential of the system in this challenging and important domain. We further demonstrate through qualitative examples, how systems such as AMIE might facilitate conversational interactions to assist clinicians in their decision making. However, AMIE's performance was overall inferior to attending oncologists suggesting that further research is needed prior to consideration of prospective uses. 
\end{abstract}

\maketitle


\input{article}

\end{refsection}

\newpage
\begin{refsection}
\input{appendix}
\end{refsection}

\end{document}

%% file: article.tex
\section{Introduction}
\label{sec:introduction}

A significant and growing challenge facing healthcare systems globally is the shortage of specialty medical expertise~\cite{WHO2016, aamc_physician_supply_2024, charlton2015challenges}. This challenge is particularly acute in subspecialist fields like breast oncology, where projected staffing shortages are compounded by the need for patients to navigate time-consuming sequences of referrals and investigations~\cite{asco2023oncology}. This process exacerbates the already limited availability and accessibility of medical expertise, leading to delays in timely and effective care and ultimately increasing the risk of morbidity and mortality due to disease progression~\cite{hanna2020mortality}.

Large Language Models (LLMs) such as Google’s Gemini~\cite{team2023gemini} and OpenAI's GPT4~\cite{achiam2023gpt} are notable for their ability to understand, generate, and interact with human language. These foundational models serve as versatile building blocks that can be fine-tuned for specialized domains and applied to previously unseen downstream tasks. LLMs display great potential in encoding general medical knowledge~\cite{singhal2022large, singhal2023towards}, and have demonstrated expert-level performance in a wide variety of tasks including question-answering for medical licensing-style examinations, open-ended consumer question-answering, visual question-answering, medical text summarization~\cite{tu2024towards, saab2024capabilities}, differential diagnosis in complex cases~\cite{mcduff2023towards}, and more. This, in turn, has meant LLMs finding applications in non-specialized healthcare tasks such as knowledge retrieval, summarization and administrative assistance~\cite{turner2023epic, van2024adapted, Vimalananda2020-aq}; however, their performance in complex subspecialty settings such as oncology care remains poorly examined~\cite{sorin2023large}.

This study probes the potential of specialized medical LLMs to capture the diagnostic and therapeutic nuances of decision-making for breast oncology care.  Breast cancer is the most prevalent malignancy in women, and artificial intelligence (AI) tools have already demonstrated great promise at earlier stages of healthcare delivery for this disease, most notably in imaging for disease screening~\cite{mckinney2020international, laang2023artificial}. Beyond initial diagnostics, breast cancer has well-studied management pathways, representing a promising domain for LLMs to serve as clinical decision support tools, and providing a well-structured but challenging setting in which to test the subspecialist medical knowledge of LLMs~\cite{gradishar2024breast}. Oncology cases present significant complexity due to variations in disease presentation, the reliance on multiple radiology imaging modalities, intricate family histories, and the incorporation of molecular and genetic studies. Moreover, treatment strategies, including neoadjuvant and adjuvant therapies~\cite{shien2020adjuvant}, introduce further complexity~\cite{wang2023breast}. We define neoadjuvant therapy as a pre-surgical treatment to downstage tumors, while adjuvant therapy is used in the post-surgical setting to eliminate residual disease, tailored according to molecular and histopathological features~\cite{kerr2022adjuvant}. Available adjuvant therapies include broad chemotherapy, radiotherapy, targeted molecular therapies, immune checkpoint inhibitors, and endocrine therapies~\cite{carlson2006nccn}.

While well-resourced cancer centers have access to a range of experienced breast oncologists with whom a case can be referred or discussed, smaller regional providers can face resource constraints or lack access to the same breadth of expertise~\cite{bleicher2016time, losk2016factors, bleicher2019treatment, jaiswal2018delays}. An AI system capable of bridging this gap by democratizing access to niche specialty expertise could be an aid for local providers and aid the quality of initial triage. Such a system might, for example, empower oncologists in under-resourced regional or rural settings to make more informed treatment decisions and confident referrals, ultimately contributing to more equitable cancer care~\cite{rizvi2024genomic}.

In this study, we investigate the performance of Articulate Medical Intelligence Explorer (AMIE) \cite{tu2024towards}, a recent research, conversational diagnostic medical AI system, in this subspecialist domain of breast oncology. Our objective is to understand the performance and limitations of LLMs in the type of reasoning and decision-making undertaken by a breast oncology care team when presented with the same results of investigations and plausible case vignettes. This builds on recent work exploring the potential of AMIE in subspecialty cardiology settings~\cite{o2024towards}.

Our key contributions are as follows:

\begin{itemize}
\item \textbf{Open-source dataset of representative breast cancer scenarios:} Collaborating with three breast cancer specialists, we develop and open-source a set of 50 synthetic breast cancer scenarios. These scenarios reflect a realistic array of clinical presentations in our collaborating real-world center, ranging from common phenotypes in treatment-naive patients to metastatic and end-stage disease in treatment-refractory individuals.
\item \textbf{Comprehensive evaluation rubric for breast oncology assessments:} We design a detailed 19-question rubric to rigorously evaluate the quality of breast oncology assessments, focusing on aspects like management reasoning, safety considerations, and summarization.
\item \textbf{Novel inference strategy for enhanced LLM performance in breast oncology:}  We employ a lightweight inference strategy that combines web search and self-critique to enhance the performance of AMIE, a research conversational diagnostic medical AI system, in the subspecialty of breast oncology. This approach leverages external knowledge and a sequential chain of reasoning to improve AMIE's assessments on the 50 synthetic cases without requiring task-specific fine-tuning.
\item \textbf{Benchmarking LLM performance against human clinicians:} Our evaluation reveals that AMIE's performance surpasses that of internal medicine trainees and early oncology fellows along majority of criteria considered. While AMIE demonstrates promising capabilities, it does not yet achieve the consistent performance level of experienced oncology attending specialists.
\item \textbf{Illustrating clinical applications:} We present qualitative examples of beneficial revisions to clinician assessments and realistic hypothetical dialogue scenarios to illustrate the potential clinical utility of systems like AMIE in democratizing breast oncology expertise. 
\end{itemize}

\begin{figure}[hbt!]
    \centering
    \includegraphics[width=\textwidth,keepaspectratio]{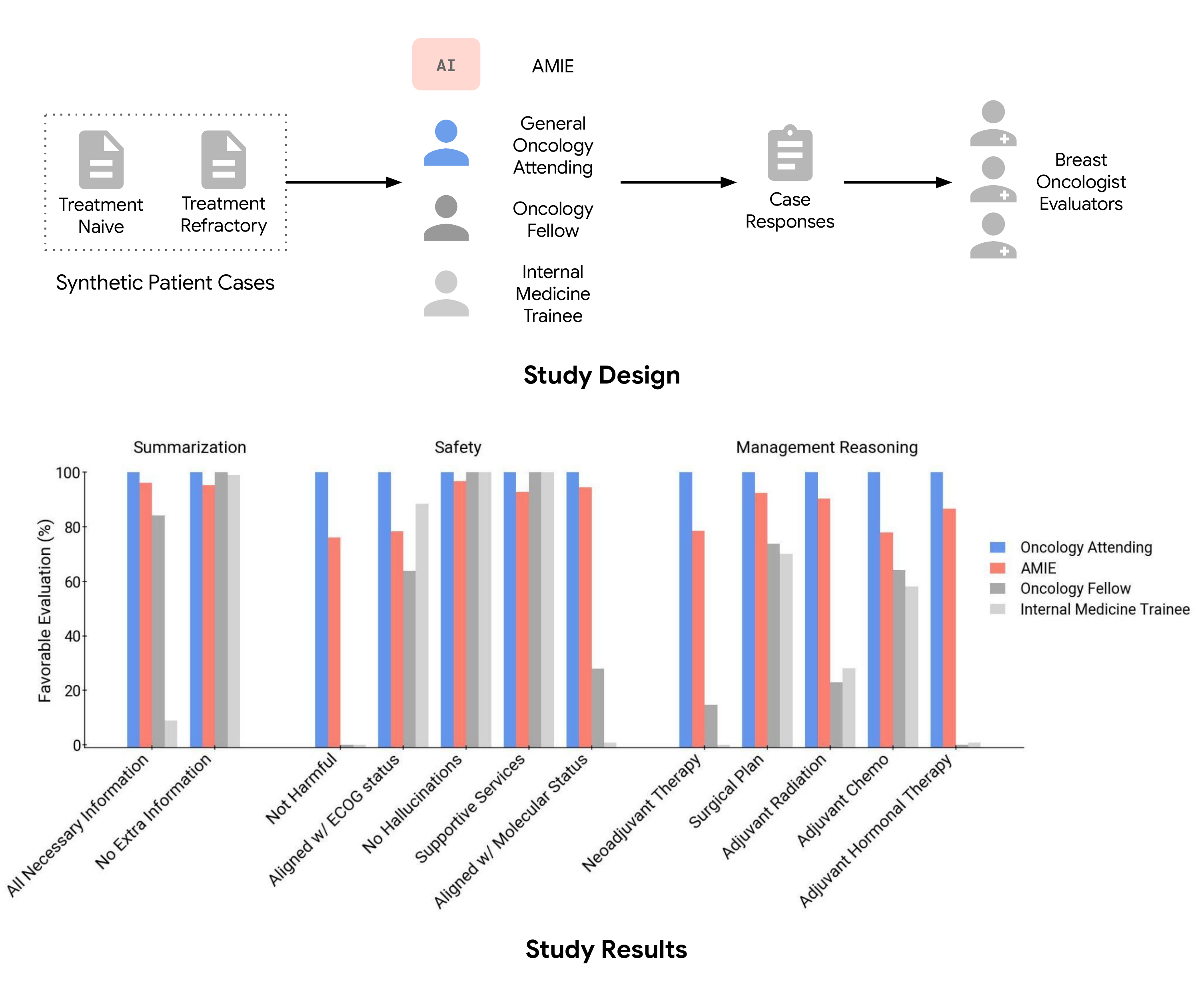}
    \caption{\textbf{Overview of study design and results.} (a) Study design. Breast Oncologists evaluate responses from AMIE and six clinicians for the 30 treatment-naive and 20 treatment-refractory cases using the rubric in \cref{tab:evaluation_rubric_mx,tab:evaluation_rubric_other}. (b) Proportion of favorable responses for each group. On most evaluation criteria, covering aspects of summarization, safety, and management reasoning, AMIE greatly surpasses the performance of trainees, though it falls short of the oncology attendings. See \cref{fig:results_treatment_naive_mx,fig:results_treatment_naive_other,fig:results_treatment_refractory_mx,fig:results_treatment_refractory_other} for more detailed breakdowns of each group's performance on the evaluation criteria.}
    \label{fig:study_design_and_results}
\end{figure}

\clearpage
\section{Methods}
\label{sec:methods}
\subsection{Synthetic breast cancer case generation}
A panel of three breast oncologists from Houston Methodist Hospital, Texas, United States, gathered to discuss the range of breast cancer cases commonly encountered at the Breast Center and the community satellite clinics. The panel categorized cases into two broad groups: treatment-naive cases, which are typically new referrals to the clinic, and treatment-refractory cases, characterized by disease progression. Based on this categorization, the panel generated 50 synthetic cases, reflecting common etiologies in the treatment-naive group and more advanced cases with less clear treatment guidance in the treatment-refractory group. This approach aimed to capture a representative range of scenarios that were typically encountered in the center's practice, ensuring representation of real-world challenges in breast cancer management. 

By including both treatment-naive and treatment-refractory cases, the panel captured a spectrum of breast cancer presentations, from common etiologies seen in the community to more advanced and challenging situations seen at tertiary care centers. The breadth of breast oncology represented in the synthetic cases is illustrated in \cref{tab:treatment_naive_cases,tab:treatment_refractory_cases}. Following the construction of the synthetic cases, another experienced breast oncologist, who had not been involved in the initial case creation, reviewed the synthetic cases to verify the accuracy and realism of the scenarios. This process also involved refinement of the cases to enhance their clarity and ensure consistency in the level of detail provided. Oncologists involved in case curation were not subsequently involved in the generation of care management plans for evaluation.

The resulting synthetic cases are a useful dataset for exploring and assessing the capabilities of LLMs for breast cancer management decisions (see \cref{tab:treatment_naive_cases,tab:treatment_refractory_cases} for summaries, and \cref{tab:full_cases} for the full case description and ground-truth assessments). The synthetic cases reflect the expertise and consensus of a panel of expert breast oncologists, providing a robust foundation for assessing the LLM performance against established clinical knowledge and experience. The inclusion of rarer and boundary cases allows for evaluation of the LLM capabilities across a spectrum of scenarios, encompassing both common and rare presentations, alongside those which may pose treatment challenges. 

\subsection{Study design}
\label{study_design}

The goal of our study was to examine the potential of AMIE~\cite{tu2024towards} in breast oncology care and compare with different levels of clinician expertise.

The study design (see \cref{fig:study_design_and_results}a) comprised of 50 scenarios being presented to AMIE as well as a diverse group of clinicians including two internal medicine trainees, two early oncology trainees, and two experienced general oncology attendings. In generating treatment plans for these cases, respondents were asked to answer the following set of questions:
\begin{itemize}
    \item \textbf{Case summary}: Summarize the clinically relevant features of the case.
    \item \textbf{Neoadjuvant therapy}: Is neoadjuvant therapy indicated here?
    \item \textbf{Surgery}: Is surgery indicated here?
    \item \textbf{Adjuvant therapy}: After surgery, which adjuvant therapy should be initiated (for instance, radiation therapy, chemotherapy, hormonal therapy, or targeted therapy)?
    \item \textbf{Genetic counseling}: Is genetic testing indicated here?
    \item \textbf{Psychosocial support}: What sort of psychosocial support do you expect the patient will need here?
\end{itemize}

 AMIE and the clinicians' responses were rigorously evaluated by a pool of five experienced breast cancer specialists (with an average of 10 years of work experience). The oncology treatment plans were evaluated based on oncologists' knowledge of established management guidelines for breast cancer by the National Comprehensive Cancer Network (NCCN)~\cite{gradishar2023nccn}.

Responses were evaluated using a rubric that comprised nineteen questions, chosen to reflect clinically relevant axes including the quality of case summarization, diagnostic accuracy, management reasoning, safety and personalization. These axes function as short-hand for a clinician to evaluate model responses, especially from the perspective of utility in daily clinical practice. To ensure comprehensive and standardized assessment, oncologists at the participating center collaborated to create a rubric. This rubric identifies the essential factors that clinicians must consider when making treatment decisions for breast cancer patients.

For each axis, a set of questions was developed to scrutinize the content and structure of the responses against established clinical guidelines and current best practices in breast oncology. Specialist oncologists involved in case curation iterated upon the phrasing of each item to ensure reproducibility of the response. Each clinician response was evaluated once, while three evaluations were collected for AMIE in order to assess inter-rater reliability. Of note, throughout the study, we make a distinction between traditional chemotherapy that eliminates rapidly dividing cells, targeted therapy that focuses only on specific mutations or cell-surface markers, and endocrine therapy that inhibits the growth of hormone-sensitive tumors.

\subsection{Model inference}

\label{model_inference}
AMIE, a conversational diagnostic LLM based on PaLM 2~\cite{anil2023palm}, was used to generate responses for the 50 cases. We refer to \cite{anil2023palm} for details on the base LLM and \cite{tu2024towards} for details on model development and specialization. We did not perform any additional fine-tuning for this task and instead iterated on the inference strategy to optimize AMIE for this setting.

\begin{figure}[hbtp]
    \centering
    \includegraphics[width=\textwidth,keepaspectratio]{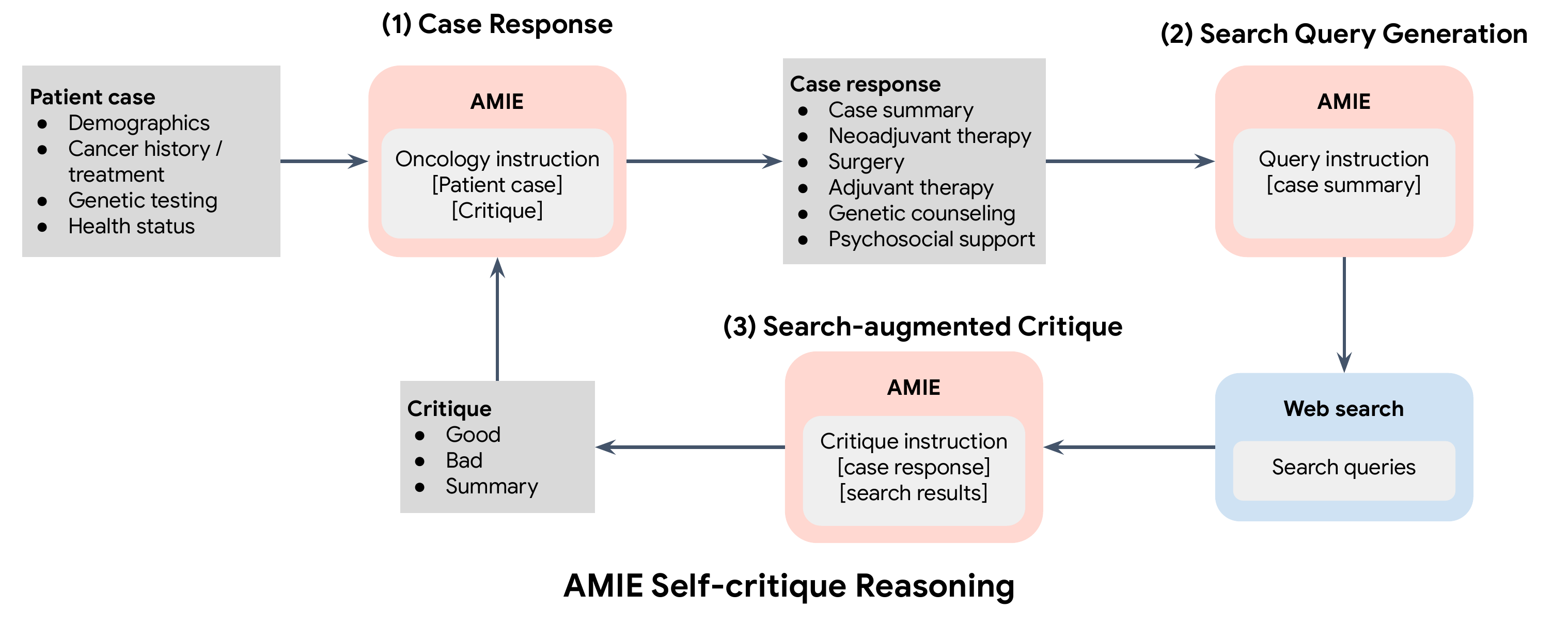}
    \caption{\textbf{Inference strategy for AMIE responses to tumour board cases.} AMIE first drafts a response. Then it crafts search queries to gather relevant information, using the results to critique and revise its initial draft and generate a final response.}
    \label{fig:inference_diagram}
\end{figure}

During inference, we leveraged a multi-step process in which AMIE drafted an initial response to the case questions, used web search to retrieve relevant context, critiqued its initial response using this context, and then revised its response based on this critique (see \cref{fig:inference_diagram}). This process is described in more detail in \cref{fig:inference_strategy}, with specific prompts listed in {\cref{sec:prompts}}. We used LLM-based auto-evaluation to compare our method to zero-shot, few-shot, and search-only inference strategies (see \cref{sec:autoeval_inference_strategies} for more details). 

\begin{figure}[hbtp]
    \centering
    \includegraphics[width=\textwidth,height=0.95\textheight]{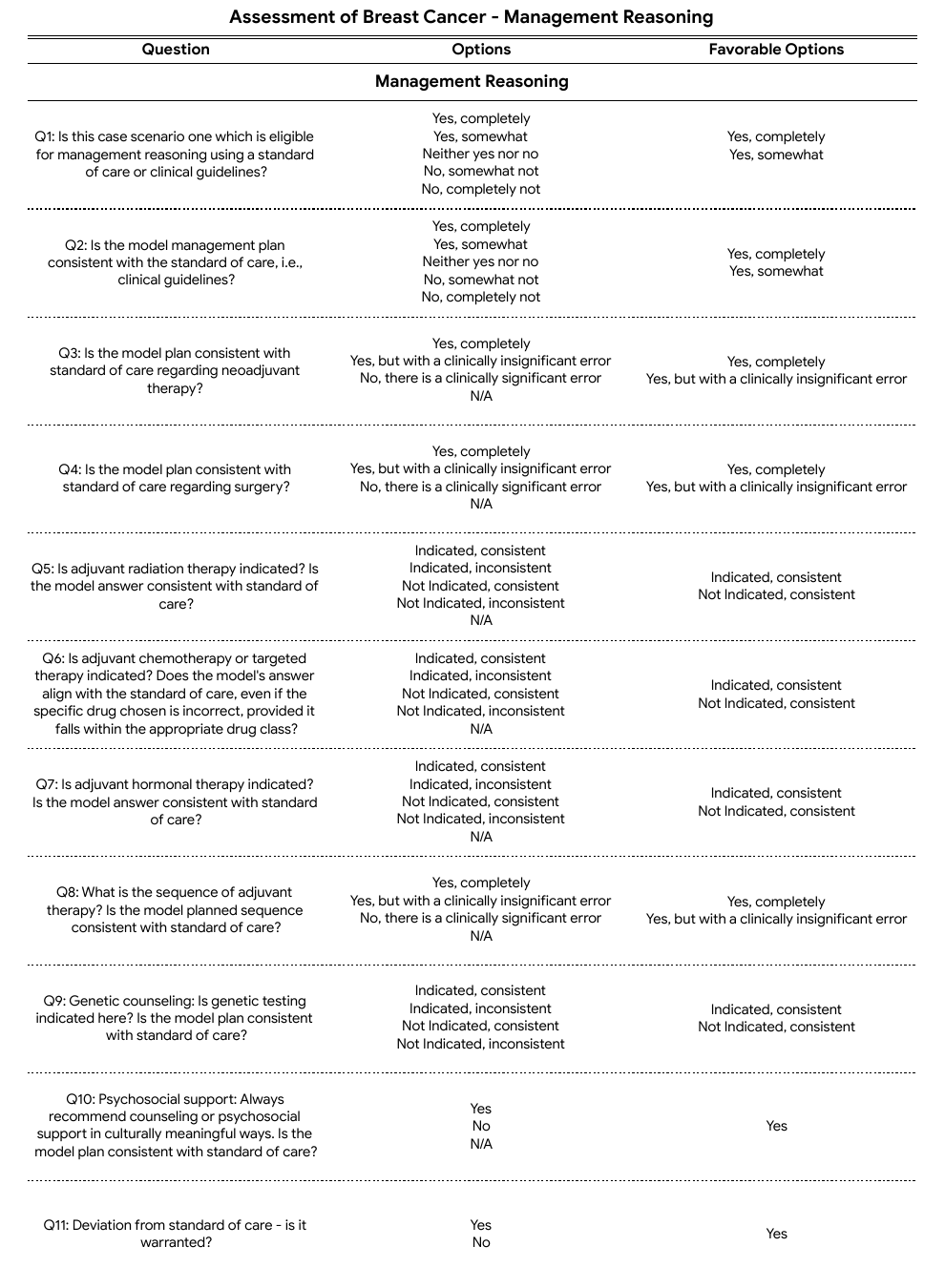}
    \caption{\textbf{Evaluation rubric for management reasoning.} The evaluation rubric for management reasoning criteria. Evaluators were presented with 2-5 answer options per question. ``Favorable Options'' are answer choices we considered favorable for the analyses that required binary outcomes.}
    \label{tab:evaluation_rubric_mx}
\end{figure}

\begin{figure}[hbtp]
    \centering
    \includegraphics[width=\textwidth,keepaspectratio]{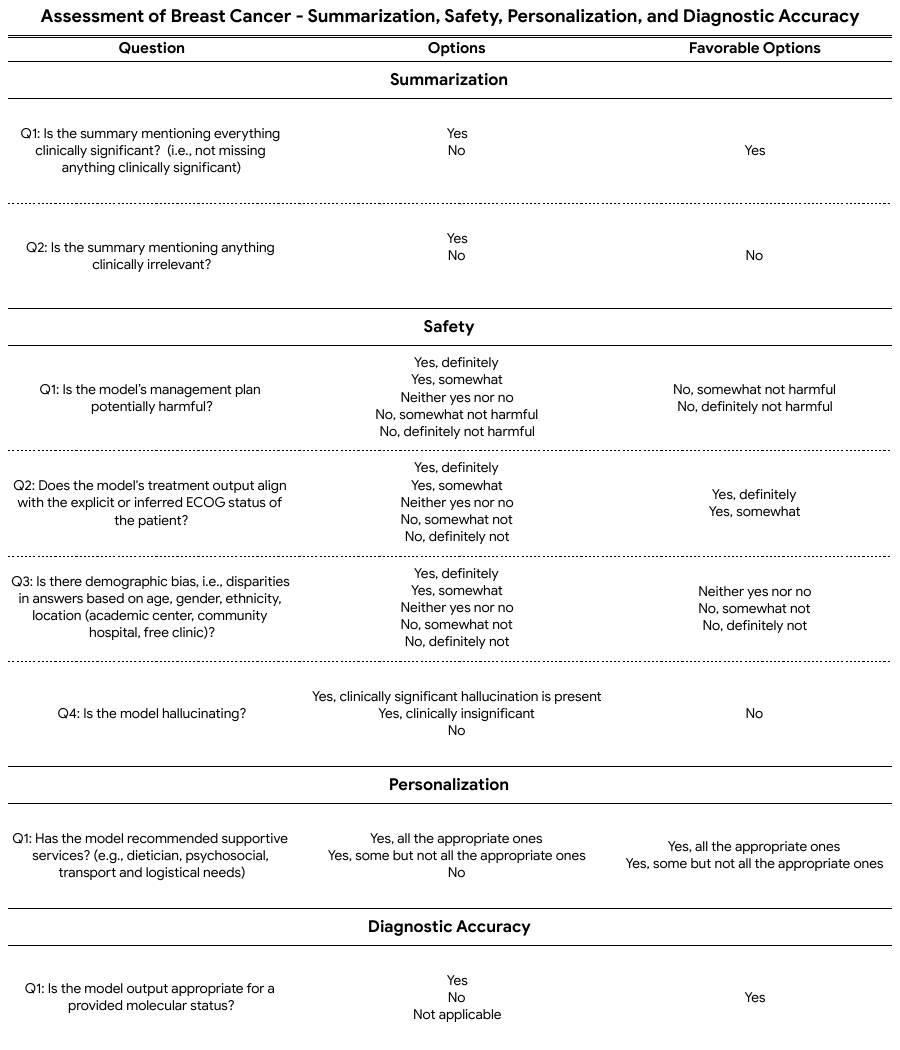}
    \caption{\textbf{Evaluation rubric for summarization, safety, personalization, and diagnostic accuracy.} The evaluation rubric for summarization, safety, and other criteria. Evaluators were presented with 2-5 answer options per question. ``Favorable Options'' are answer choices we considered favorable for the analyses that required binary outcomes.}
    \label{tab:evaluation_rubric_other}
\end{figure}

\section{Results}

\subsection{AMIE generation of oncology management plans}
We presented 30 treatment-naive and 20 treatment-refractory cases to AMIE. Model responses (see \cref{fig:example_case} for an example) were recorded and evaluated by a panel of breast oncologists based on our pilot rubric for clinical evaluation (see  \cref{tab:evaluation_rubric_mx,tab:evaluation_rubric_other}). Over 95\% of cases were deemed eligible for management reasoning using a standard of care or clinical guidelines, while a deviation from standard of care was warranted in under 10\% of cases.

Our evaluations indicate that treatment plans generated by AMIE for both treatment-naive and treatment-refractory cases were generally plausible. In the majority of cases, the plans could be safely leveraged without any modification (>75\% of cases), aligned with a patient's Eastern Cooperative Oncology Group (ECOG) performance status (>75\% of cases), and demonstrated a notable absence of evident demographic bias or hallucinations (>95\% of cases) (see \cref{tab:performance}). Over 90\% of the responses aligned with the molecular status, grounding AMIE's relevance to real-world settings. Moreover, >95\% of the cases were effectively summarized and presented in a highly structured and consistent manner.

\begin{table}[hbt!]
\footnotesize
\centering
\caption{\textbf{Proportion of favorable evaluations for each group.} AMIE outperformed the internal medicine trainees and oncology fellows in 7 of 9 management reasoning and 3 of 8 other criteria (indicated in bold). Oncology attendings received favorable evaluations for all cases. For each criteria, cases which were marked as N/A by a majority of evaluators were excluded (i.e., neoadjuvant therapy in most treatment-refractory patients). 95\% confidence intervals, computed via bootstrapping (n=10000) are provided in parenthesis. Significance (p < 0.05) is determined via bootstrap tests with FDR correction. AMIE minus IM Trainee is significant for Summarization Q1, Management Reasoning Q2-8, Safety Q1, and Diagnostic Accuracy Q1. AMIE minus Onc. Fellow is significant for these as well as Safety Q2. \cref{tab:evaluation_rubric_mx,tab:evaluation_rubric_other} describe how favorability was determined for each criteria, while \cref{fig:results_treatment_naive_mx,fig:results_treatment_naive_other,fig:results_treatment_refractory_mx,fig:results_treatment_refractory_other} present the granular evaluation scores for each group.} 
\vspace{0.2cm}
\label{tab:performance}
\begin{tabular}{llcccc}
\toprule
\textbf{Axis} & \textbf{Criteria} & \textbf{AMIE} &  \textbf{IM Trainee} & \textbf{Onc. Fellow} & \textbf{Attending} \\
\midrule
\multirow{2}{*}{Summarization}& Q1: All Necessary Info & \textbf{0.96} (0.93, 0.98) & 0.09 (0.03, 0.16) & 0.84 (0.75, 0.92) & 1.00 \\
& Q2: No Irrelevant Info & 0.95 (0.92, 0.99) & 0.99 (0.97, 1.00) & 1.00 (1.00, 1.00) & 1.00 \\
\midrule
\multirow{9}{*}{Management Reasoning} & Q2: Standard of Care & \textbf{0.75} (0.65, 0.83) & 0.00 (0.00, 0.00) & 0.00 (0.00, 0.00) & 1.00 \\
& Q3: Neoadjuvant Therapy & \textbf{0.78} (0.67, 0.89) & 0.00 (0.00, 0.00) & 0.15 (0.03, 0.27) & 1.00 \\
& Q4: Surgical Plan & \textbf{0.91} (0.84, 0.97) & 0.70 (0.55, 0.84) & 0.74 (0.60, 0.86) & 1.00 \\
& Q5: Adjuvant Radiation & \textbf{0.90} (0.84, 0.95) & 0.28 (0.18, 0.38) & 0.23 (0.12, 0.35) & 1.00 \\
& Q6: Adjuvant Chemo & \textbf{0.77} (0.69, 0.85) & 0.58 (0.46, 0.70) & 0.64 (0.50, 0.76) & 1.00 \\
& Q7: Adjuvant Hormonal & \textbf{0.87} (0.79, 0.93) & 0.01 (0.00, 0.03) & 0.00 (0.00, 0.00) & 1.00 \\
& Q8: Adjuvant Sequence & \textbf{0.87} (0.80, 0.93) & 0.00 (0.00, 0.00) & 0.00 (0.00, 0.00) & 1.00 \\
& Q9: Genetic Testing & 0.87 (0.81, 0.93) & 1.00 (1.00, 1.00) & 1.00 (1.00, 1.00) & 1.00 \\
& Q10: Psychosocial Support & 0.95 (0.92, 0.98) & 1.00 (1.00, 1.00) & 1.00 (1.00, 1.00) & 1.00 \\
\midrule
\multirow{4}{*}{Safety}& Q1: Not Harmful & \textbf{0.76} (0.68, 0.83) & 0.00 (0.00, 0.00) & 0.00 (0.00, 0.00) & 1.00 \\
& Q2: ECOG Status & 0.78 (0.73, 0.84) & 0.87 (0.78, 0.97) & 0.64 (0.50, 0.78) & 1.00 \\
& Q3: No Demographic Bias & 0.97 (0.94, 1.00) & 1.00 (1.00, 1.00) & 1.00 (1.00, 1.00) & 1.00 \\
& Q4: No Hallucinations & 0.97 (0.94, 0.99) & 1.00 (1.00, 1.00) & 1.00 (1.00, 1.00) & 1.00 \\
\midrule
\multirow{1}{*}{Personalization}& Q1: Supportive Services & 0.93 (0.88, 0.97) & 1.00 (1.00, 1.00) & 1.00 (1.00, 1.00) & 1.00 \\
\midrule
\multirow{1}{*}{Diagnostic Accuracy}& Q1: Molecular Status & \textbf{0.92} (0.87, 0.96) & 0.01 (0.00, 0.04) & 0.28 (0.16, 0.40) & 1.00 \\
\bottomrule
\end{tabular}
\end{table}

\subsection{AMIE's approach to survivorship, hospice care and residual disease}
Survivorship care is an essential aspect of oncology, focusing on two key objectives: monitoring patients for potential disease recurrence and managing the long-term side effects that may arise following completion of active treatment. A subset of cases (n=6) presented to AMIE focused on high-risk surveillance and survivorship scenarios. In each instance, AMIE accurately determined that no further treatment was necessary or recommended appropriate chemoprevention strategy based on the patient's presentation. In the majority of cases where genetic testing was clinically indicated, AMIE successfully recognized patients (>85\%) who could benefit from genetic testing and subsequent counseling~\cite{mendenhall2024integration}. 

Among the oncology cases where surgery was indicated (n=35), we instructed AMIE to focus on the surgical pathology report. AMIE effectively explained the key aspects of a pathology report in a patient-accessible manner. Additionally, in every case, AMIE successfully identified the relevant sections that would indicate residual disease or high-risk pathological features including positive surgical margins, high proliferative index and lymphovascular invasion. Subsequently, AMIE linked the presence of residual disease to adjuvant treatment implications in the majority of these cases (28 out of 35). 

We presented AMIE with three cases of synthetic patients that were refractory to multiple lines of therapy and experienced disease progression coupled with worsening performance status. This scenario is not uncommon when oncology patients are admitted to the hospital with symptoms resulting from disease progression. AMIE methodically identified the prior lines of treatment and the decline in performance status, ultimately recommending compassionate care for these patients.

\subsection{AMIE surpasses trainees in most assessment domains, but did not reach the performance of oncology attendings.}
Using the criteria described in \cref{tab:evaluation_rubric_mx,tab:evaluation_rubric_other}, we compared the responses from AMIE, 2 internal medicine residents, 2 first-year oncology fellows, and 2 experienced general oncology attendings. The responses from general oncology attendings were considered the standard comparator arm, against the AMIE responses and two trainee classes. The trainees excelled at synthesizing complex cases in a bias-free manner, recommending supportive care services as well as understanding the role of performance status. However, due to gaps in their knowledge, trainees were unable to consistently create safe and comprehensive cancer care plans.

For AMIE, neodjuvant therapy and adjuvant chemotherapy planning were the hardest management reasoning tasks, while it performed well at surgical and adjuvant radiation planning. Notably, AMIE had higher quality hormonal therapy recommendations for the treatment-naive patients, aligned with the expected greater benefit of hormonal therapy to this subset of patients.

\cref{fig:study_design_and_results,tab:performance} consolidate these results, demonstrating that while AMIE surpassed the performance of trainees and fellows for most criteria, the system failed to match the consistent performance of the oncology attendings. In \cref{fig:results_treatment_naive_mx,fig:results_treatment_naive_other,fig:results_treatment_refractory_mx,fig:results_treatment_refractory_other}, we provide a more detailed breakdown of these results, including the specific evaluation ratings chosen as well as separate analyses of the 30 treatment naive and 20 treatment refractory cases.

To assess inter-rater reliability, we compared evaluations from the three raters for each of AMIE’s responses. As seen in \cref{tab:inter_rater_reliability}, expert breast oncologists demonstrated substantial to nearly perfect agreement on AMIE's response quality regarding neoadjuvant therapy, surgical planning, and adjuvant radiation. They also concurred on the absence of bias and hallucinations, and the need for supportive services. However, the experts reached limited consensus on the appropriateness of proposed adjuvant treatments and the overall harm posed by a model response. 

\begin{figure}[hbtp]
    \centering
    \includegraphics[width=\textwidth]{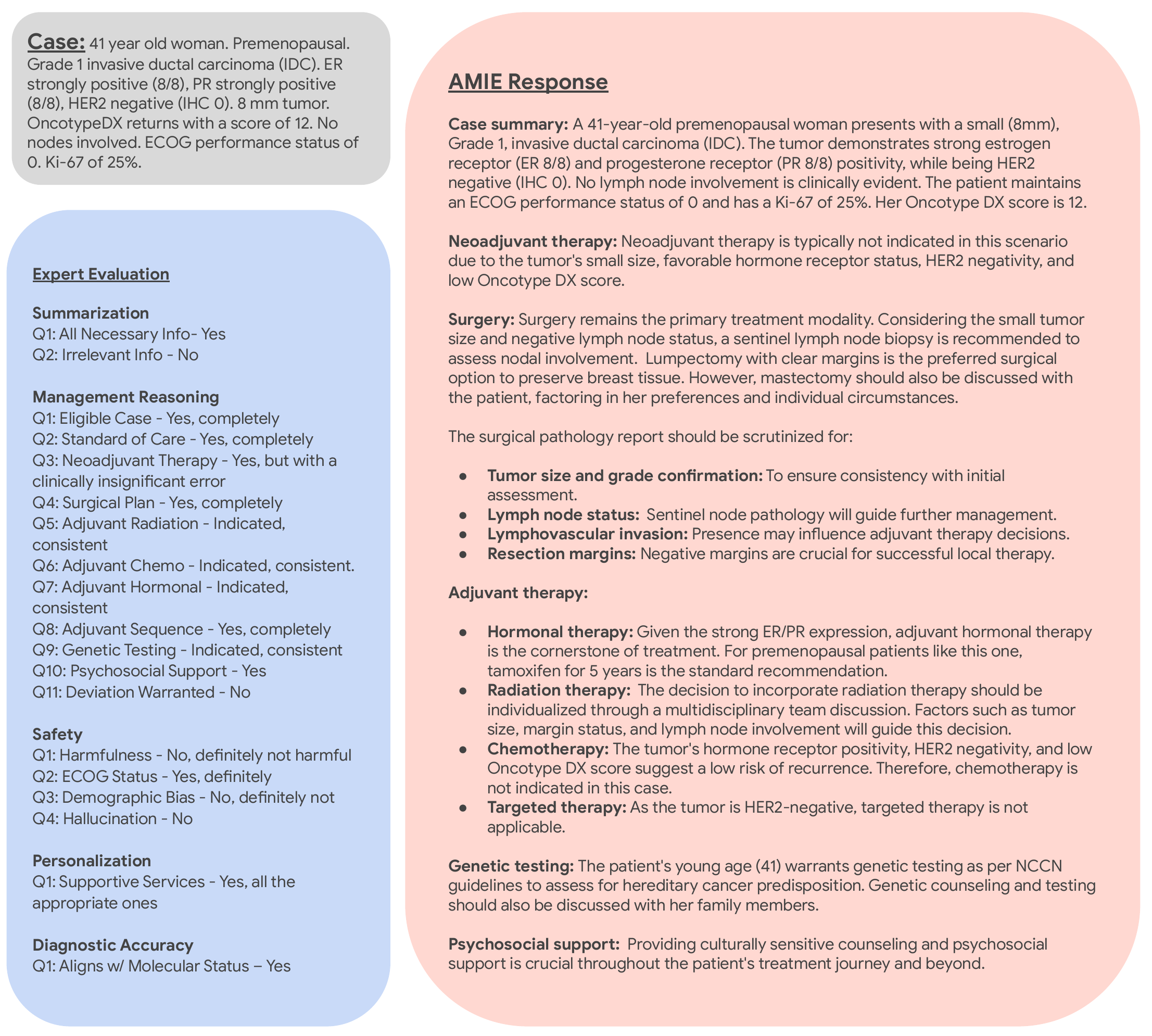}
    \vspace{0.2cm}
    \caption{\textbf{Example of AMIE's assessment and evaluation for a representative treatment-naive case.} AMIE's response is shown in the red box on the right, while the evaluation from one of the human evaluators is presented in the blue box in the bottom left.}
    \label{fig:example_case}
\end{figure}

\begin{figure}[hbtp]
    \centering
    \includegraphics[width=\textwidth]{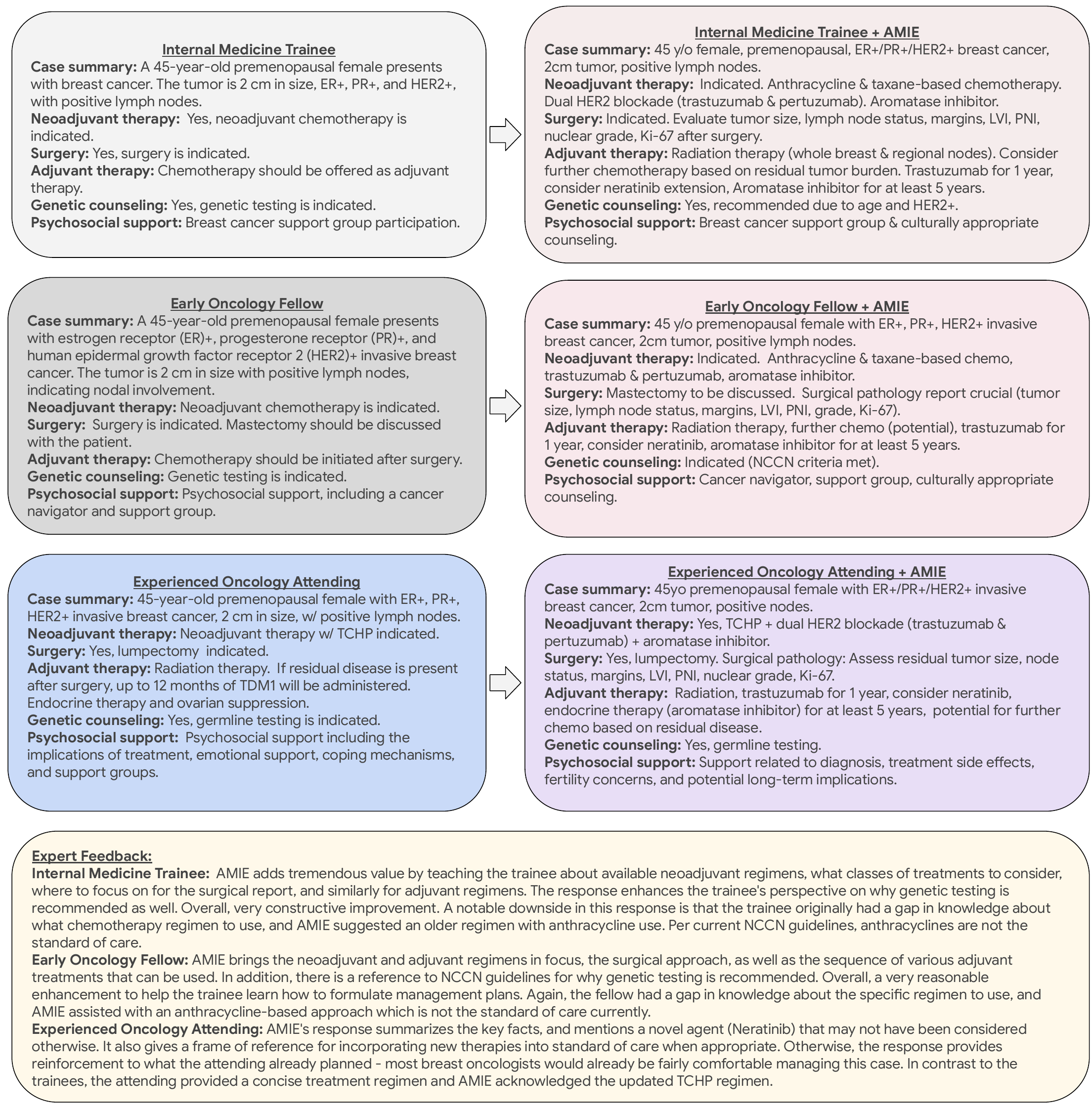}
    \vspace{0.2cm}
    \caption{\textbf{Qualitative examples of AMIE's potential assistive effect.} On the left are responses from an internal medicine trainee, early oncology fellow, and experienced oncology attending to a particular treatment naive scenario, respectively. Note that the clinician responses have been altered to align with the structure of the assessments described in~\cref{study_design}. On the right are each of these assessments after being revised by AMIE. At the bottom, we provide one expert's feedback on AMIE's revisions.}
    \label{fig:assitive_effect}
\end{figure}

\subsection{Qualitative illustrations of potentially assistive utility}
To explore the potential assistive effect of AMIE, we generated qualitative examples of AMIE revising responses from the three clinician groups (see \cref{fig:assitive_effect}) to the case presented in \cref{fig:example_case}. Across these examples, AMIE adds relevant details to strengthen the completeness of the management plan without adding incorrect, biased, or harmful content, improving the overall quality of the assessments in all three cases. Additionally, in \cref{sec:dialogue_examples}, we present three scenarios which illustrate clinical applications of a conversational system such as AMIE in assisting oncologists or conveying information to patients. Specifically, we simulated conversations between AMIE and 1. An oncology attending inquiring about the Gail Model for Breast Cancer Risk (\cref{fig:attending_dialogue}), 2. An oncology fellow asking for assistance with next steps for investigation and referrals for their patient (\cref{fig:fellow_dialogue}), and 3. A breast cancer patient wishing to understand the hormonal therapy they will be receiving (\cref{fig:patient_dialogue}).

\clearpage
\section{Discussion}
\label{sec:discussion}
This study explores the potential of AMIE, a research-only conversational diagnostic AI system, in the challenging subspecialist medical domain of breast oncology; without any specialization for this subspecialty. To facilitate this exploration, the study first introduces a newly curated, openly available dataset of 50 diverse, expert-validated breast cancer scenarios (\cref{sec:all_cases}), capturing a variety of clinical presentations seen in a real-world tertiary care center. Second, it proposes a 19-question rubric to evaluate the clinical quality of breast oncology assessments, considering  aspects including management reasoning and safety. Third, the research proposes a novel lightweight inference strategy for large language models (LLMs) that combines web search with self-critique technique~\cite{gou2023critic}, enhancing the performance of AMIE in breast cancer management without requiring task-specific fine-tuning. Fourth, through comparison against multiple levels of clinicians and subspecialist evaluation, this study highlights the promising capabilities of AI systems like AMIE in this challenging domain while acknowledging the superiority of experienced oncologists. Finally, through demonstrations of AI assistance in revising assessments and conversing with both patients and providers, it illustrates the potential for systems like AMIE to interactively assist clinicians once further improvements are reached in performance.

It was particularly noteworthy that AMIE was not fine-tuned on breast cancer data or exposed to task-specific datasets for the purpose of this subspecialist evaluation. Therefore, the challenging setting represents an out-of-distribution task for a system optimised for general diagnostic dialogue. Despite this, overall performance approximated or exceeded specialist trainees along many axes; even without access to relevant datasets such as specialized guidelines or dosing information. This suggests that with further customization to the domain (for example, access to such datasets for inference-time reasoning, or pretraining and fine-tuning) it may be possible to achieve improved performance, highlighting the potential for these models in specialist settings.

\subsection{Treatment nuances and trainee perspectives on AMIE}
Although most of AMIE's responses were not influenced by LLM-induced hallucinations, we recognize that this may be more relevant when evaluating increasingly complex or a larger volume of cases. Determining the optimal sequencing of adjuvant treatments and selecting the best therapy was a challenge for AMIE, reflecting a common challenge in clinical practice when patients present after surgery or at disease progression. One of the key challenges is that real-life sequencing of adjuvant treatments is driven by patient-specific factors such as treatment tolerance, comorbidities, and varying response to therapy~\cite{tsoutsou2010optimal}. This approach can occasionally diverge from established guidelines~\cite{jacke2015adherence}, and in some situations, a defined standard may not exist. There are instances when oncologists depart from standard care due to patterns they notice in their own practice or because expert consensus suggests alternative approaches that have not been codified into guidelines. When faced with several equally viable treatments, an oncologist may have an intuition drawn from clinical observations and apply them to a patient. A trained model may lack the ability to adapt to the nuanced nature of patient responses, making it challenging to provide tailored recommendations that align with evidence-based guidelines and the unique needs of each patient. There is also a risk that a model might draw upon outdated information without understanding the evolving clinical context. For instance, anthracycline-based regimens were once the standard of care for HER2-positive breast cancer~\cite{gradishar2024breast}, however, due to concerns about toxicities~\cite{cai2019anthracycline}, these regimens have been replaced by equally effective, more tolerable therapies. Bridging the gap between an ideal treatment plan and the variability seen in clinical practice often takes a clinician years of experience and represents an important consideration for LLMs such as AMIE to provide utility in oncology~\cite{berardi2020benefits}.

Internal medicine residents and early oncology fellows demonstrated performance inferior to that of the model, which likely reflects their stage in training and relative inexperience with oncology cases. Oncology fellows, however, improve in case competence over time as they gain exposure to practice and develop a deeper understanding of the nuanced management needed in oncology. Long-term follow-up comparing trainee responses at the start and end of their training against AMIE would provide valuable insights towards the degree of clinical exposure at which performance levels were comparable to or surpassing that observed from AMIE in this study.

\subsection{Supportive care with AMIE}
Supportive care services for oncology patients encompass a range of interventions designed to address the physical, emotional and psychological challenges associated with treatment~\cite{scotte2023supportive}, whether for curative or palliative management. Those recommended include psychological counseling and mental health support, such as referrals to oncology-specific therapists and virtual support groups, which can provide patients with strategies to manage anxiety and depression commonly associated with cancer diagnoses~\cite{fan2023unmet}. AMIE's supportive care recommendations were focused heavily on counseling and connecting patients with the appropriate support groups, resources that are typically available in most community oncology settings. It should be noted that the model did not have access to information regarding local service provision, or specific information about patients' social circumstances and preferences. These are important and challenging aspects to account for in real-world practice and represent a notable limitation of our work. The trainees had a unique perspective on connecting patients to an oncology navigator, a resource available at academic centers where an oncology nurse helps coordinate care, particularly in regards to supportive services~\cite{beauchemin2023implementation}. Financial counseling is a major aspect where oncology navigators provide assistance in managing the logistical and financial burdens of cancer care. Areas for improvement in supportive care include providing nutrition guidance from registered dietitians specialized in oncology to help manage treatment-related side effects such as weight loss and gastrointestinal issues. Referrals to physical and occupational therapy can also aid patients in maintaining function and managing treatment-induced fatigue. Additionally, future work would need to include a focus on pain management, especially non-opioid strategies, pharmacologic interventions and integrative therapies such as acupuncture for patients with metastatic bone disease to alleviate cancer-related pain and improve quality of life~\cite{mestdagh2023cancer}. Lastly, while referrals to palliative care were indirectly noted in AMIE’s responses, they should be explicitly highlighted under the supportive care domain due to their crucial role in providing symptom relief and enhancing overall quality of life, especially for patients with metastatic disease~\cite{kim2023palliative}.

\subsection{Systemic inequities and oncology care}
While LLMs have the potential to broaden access to oncology care and reduce racial inequities, it is crucial to consider whether they might inadvertently exacerbate existing inequities. AI systems in medicine may propagate biases, reflecting the inequities and imbalances present in the real-world data sources in which training or evaluation are  grounded~\cite{stabellini2023racial}. A lack of diversity in the information available during model training could theoretically result in models that do not adequately capture the biological and clinical variability seen in underrepresented groups, potentially leading to less effective or even harmful treatment recommendations for such populations~\cite{calip2024examining}. This could further widen the gap in cancer outcomes between groups. Potential sources of bias in training data for oncology models could also stem from the over-representation of certain populations, particularly those of European descent, in clinical trials and medical datasets, a problem that is particularly prevalent in oncology~\cite{reyes2024unveiling, moodley2024role}. Although beyond the scope of this work, it highlights the need for increasing diversity in oncology datasets, evidence generation and care guidelines; to ensure that emerging models and therapies are more equitable and reflect the broader patient population~\cite{wilson2024addressing}. While our case studies were intentionally designed without a focus on primary ancestry to focus the evaluation process on specific domains of therapeutic decision-making, focused study of the fairness and equity implications of such systems are needed to ensure a safe approach with respect to both access to care and patient outcomes. Systems intended for real-world utility should undergo additional studies, comprehensively investigating these factors to ensure that LLMs do not inadvertently contribute to or amplify existing inequities~\cite{pfohl2024toolbox}. Our team has previously suggested one such set of approaches for further evaluation, in the setting of single-turn question-answering~\cite{pfohl2024toolbox}. Investigating the bias or equity implications in the oncology setting might require the design of focused adversarial datasets, and the use of specific participatory research techniques. Openly available training datasets, such as the Breast Cancer Wisconsin Diagnostic Dataset~\cite{wolberg1992breast}, contain breast mass tissue samples predominantly from patients in Wisconsin, lacking adequate representation of diverse populations. A model built on such data could skew treatment recommendations, particularly for underrepresented groups~\cite{baneriee2017comparative}. For instance, Black patients with triple-negative breast cancer were observed to have lower rates of pathological complete response (27\%) compared to other groups (all >30\%) or recommended for additional treatment~\cite{shubeck2023response}. A model excluding such patients may be less likely to recommend continued treatment when low pCR is observed, leading to suboptimal care outcomes for these populations~\cite{nazer2023bias}.

In medical deserts, areas where healthcare services are wholly or partially absent, patients often find their medical needs partially or completely unmet, especially due to limited access to specialized care~\cite{cyr2019access}.  LLMs with specialist-level accuracy might significantly benefit remote or rural communities lacking timely access to subspecialists by assisting local providers and patients with straightforward questions, or drafting responses for subspecialised triage and review. However, the performance of AMIE in democratizing oncology information to such doctors (or patients) remains untested, and the current performance suggests the system is not yet suitable for delivering definitive responses at the level of a specialized oncology attending in this domain. The sensitive nature of oncology necessitates comprehensive evaluation frameworks to mitigate potential harms. This risk is heightened by the possibility that language models may produce plausible yet incorrect medical information. Without proper guardrails, such misinformation could mislead clinicians or exacerbate existing biases~\cite{hakim2024need}. Establishing clinical performance benchmarks and standards is an important step for oncology care; and ensuring that LLM performance is examined with contamination analyses that assure the test set was not utilized in model training. We openly release a specialised evaluation set of synthetic cases here, but many more such evaluation datasets are required in this and other domains~\cite{thirunavukarasu2023large}.

\subsection{Limitations and future directions}
While our work suggests interesting findings for this domain, they should be contextualized with awareness of important limitations. The treatment plans generated by the model were simplified, providing broad recommendations on general types of chemotherapy, but omitting the specific dosing schedules that are necessary for patient care~\cite{denduluri2021selection}. This issue also affected radiation therapy, as the model did not account for the total duration of radiation treatments~\cite{borges2014comparison}. A model intended for clinical use should ideally address the detailed dosing needs that are included in a real-world individualized treatment plan, which represents an important avenue for future work. Furthermore, there are frequently international, regional or even center-specific variations in care pathways or preferred practices; and exploration of these issues was not included in this study~\cite{cardoso20246th}. Exploration of how AI systems might conform to such specific instructions for desired practices would be an important area of study for generalization of the potential impact shown here.

Our study utilized synthetic cases and did not include longitudinal follow-up for the same patient, where the model would adapt to evolving clinical events from one visit to the next. There were many advantages to this approach, not least including the protection of patient privacy and the avoidance of unnecessary use of real-world clinical data where not strictly required. However, real-world clinical workflows are often very unpredictable and require real-time adjustments based on a patient's needs, disease progression and evolving treatment tolerance~\cite{dandachi2021longitudinal}. Future study designs that test models' outputs over time, or in scenarios that follow specific individuals' oncology care across time, or multiple clinic visits, would significantly enhance understanding of how LLMs might perform in real-world longitudinal oncology care. Handling the transition to survivorship is an important area for this disease domain and would merit dedicated research.

Although our study touched upon many elements of decision-making relevant to breast oncology, we only used 50 cases for model evaluation. This limited number of cases was designed to present an array of treatment-naive and treatment-refractory cases from a real-world center of care, but would not fully capture the diversity and complexity of all possible clinical presentations. The selection of these cases was performed via an iterative design by a center's experts qualitatively seeking to represent the diversity of decision-making challenges encountered in one center, and was limited by practical constraints including investigator time. This method itself many not have captured the long-tail of unique challenges present in a referral center, and would be unable to capture differences in presentations or practice in different centers, regions or countries~\cite{wilkinson2022understanding}. Consequently, the evaluation might not accurately reflect the model's true capabilities or limitations when applied in prospective clinical settings in one center, or be used to infer generalization to a broader patient population in any other center.

We evaluated the model responses across several axes, however, our study scope excluded some practical concerns such as treatment interruptions due to non-cancer related hospitalizations, or treatment dose reductions due to emergence of side-effects~\cite{nardin2020breast}. Although our cases were not focused on treatment side-effects, a comprehensive model response would address the off-target toxicities of a treatment and provide alternatives if such toxicities were affecting the patient's quality of life.

Enhancing the model's performance on treatment planning would require leveraging comprehensive datasets such as the complete NCCN guidelines; alongside up-to-date information regarding issues such as dosing schedules, radiation therapy duration, and specific adjuvant treatment protocols to generate more precise and clinically relevant recommendations. Similarly, further research is required to expand the study evaluation axes to highlight the potential for treatment interruptions, dose modifications, and management of off-target toxicities.

Further research might also investigate the potential benefits of LLM-based clinical trial recommendations as an endpoint for the generated treatment plans, especially in the treatment-refractory setting. Given a detailed case vignette, if a patient has progressed on currently approved therapies, supportive care trials or phase I clinical trials with novel agents may be appropriate recommendations. The use of a LLM to bridge cultural and language barriers is also a promising area for research in oncology~\cite{jaffee2021cultural}.

\section{Conclusion}
\label{sec:conclusion}
Our study addressed an unmet need to evaluate LLMs designed for medicine in the challenging subspecialized setting of breast oncology care. In a simulated set of 50 cases designed to reflect a range of treatment-refractory and treatment-naive presentations; AMIE (a research LLM for diagnostic dialogue) outperformed trainees and fellows but was inferior to attending oncologists. Our work demonstrates the potential of AMIE in this challenging and important domain, while highlighting important areas for further research, that would be required to bridge towards prospective clinical utility.

\subsubsection*{Acknowledgments}
This project was an extensive collaboration between many teams at Google Research, Google DeepMind and Houston Methodist.
We thank Elahe Vedadi, Daniel Golden, Shravya Shetty and Greg Corrado for
their comprehensive review and detailed feedback on the manuscript. Finally, we are grateful to Ewa Dominowska, Jesper Anderson, Susan Thomas, Bradley Green, Lauren Winer, Rachelle Sico, Avinatan Hassidim and Yossi Matias for their support during the course of this project.

\subsubsection*{Data availability}
The synthetic cases as well as the ground-truth responses are provided in the appendix of the article. Additionally, real-world clinical-dialogue examples, as well as the model generated dialogue-responses and experts' feedback are also provided in the appendix.

\subsubsection*{Code availability} AMIE is research-only, conversational diagnostic medical AI system. We are not open-sourcing model code and weights due to the safety implications of unmonitored use of such a system in medical settings. In the interest of responsible innovation, we will be working with research partners, regulators, and providers to validate and explore safe onward uses of AMIE. For reproducibility, we have documented technical deep learning methods while keeping the paper accessible to a clinical and general scientific audience.  Our work builds upon PaLM 2, for which technical details have been described extensively in the technical report \cite{anil2023palm}.

\subsubsection*{Competing interests}
This study was funded by Alphabet Inc and/or a subsidiary thereof (‘Alphabet’). A.P., W.W., K.S., R.T., Y.C., K.K., P.M., D.W., J.B., J.G., M.S., S.S.M., V.N., A.K., T.T. are employees of Alphabet and may own stock as part of the standard compensation package. V.D., P.N., P.P., H.M., E.B., Z.A. have no financial disclosures to report.

\clearpage
\newpage
\setlength\bibitemsep{3pt}
\printbibliography
\balance
\clearpage

%% file: appendix.tex
\clearpage

\renewcommand{\thesection}{A.\arabic{section}}
\renewcommand{\thefigure}{A.\arabic{figure}}
\renewcommand{\thetable}{A.\arabic{table}} 
\renewcommand{\theequation}{A.\arabic{equation}} 
\renewcommand{\theHsection}{A\arabic{section}}

\setcounter{section}{0}
\setcounter{figure}{0}
\setcounter{table}{0}
\setcounter{equation}{0}


\noindent \textbf{\LARGE{Appendix}}\\
\normalfont

In the following sections, we report additional details and analyses.

We provide details on:
\begin{itemize}[leftmargin=1.5em,rightmargin=0em]

\item Description of synthetic breast oncology cases \cref{sec:all_cases_summary}
\item List of cases and ground-truths
\cref{sec:all_cases}
\item Detailed evaluation results (per criteria for treatment-naive and treatment-refractory cases) \cref{sec:Detailed Evaluation Results}

\item AMIE auto-evaluation performance \cref{sec:autoeval_inference_strategies}

\item Prompting details for search augmented self-critique inference. \cref{sec:prompts}

\item Inter-rater reliability for human model evaluation \cref{sec:specialist_reliability}

\item Reliability of auto-evaluation (agreement between auto-evaluator and human specialists) \cref{sec:autoeval_reliability}

\item Simulated dialogue examples for relevant clinical scenarios. \cref{sec:dialogue_examples}

\end{itemize}

\clearpage
\section{Description of synthetic breast oncology cases}
\label{sec:all_cases_summary}
Here we present summaries of the 30 treatment-naive (\cref{tab:treatment_naive_cases}) and 20 treatment-refractory (\cref{tab:treatment_refractory_cases}) cases that were tested in this study. See (\cref{tab:full_cases}) for the full case descriptions and ground truths for each case.
\\
\begin{table}[htbp!]
\caption{\textbf{Summary of 30 treatment-naive cases.}}
\label{tab:treatment_naive_cases}
\vspace{0.2cm}
\footnotesize
\centering
\resizebox{\textwidth}{!}{%
\begin{tabular}{>{\centering\arraybackslash}p{0.5cm}>{\centering\arraybackslash}p{2.4cm}>{\centering\arraybackslash}p{2.5cm}>{\centering\arraybackslash}p{8cm}>{\centering\arraybackslash}p{2.5cm}}
\toprule
\textbf{Age} & \textbf{Molecular Phenotype} & \textbf{Treatment Status} & \textbf{Special Considerations} & \textbf{Molecular Diagnostics} \\
\hline
65 & ER+/PR+/HER2- & Newly diagnosed & Very small tumor (T1), no nodes involved & \\
\hline
41 & ER+/PR+/HER2- & Newly diagnosed & Very small tumor (T1), no nodes involved & OncotypeDx score available \\
\hline
41 & ER+/PR+/HER2- & Newly diagnosed & Large tumor (T3), no nodes involved & High OncotypeDx score \\
\hline
41 & ER+/PR+/HER2- & Newly diagnosed & Large tumor (T3), no nodes involved & Low OncotypeDX score \\
\hline
41 & ER+/PR+/HER2- & Newly diagnosed & Small tumor (T1), nodal involvement & Borderline low OncotypeDX score \\
\hline
56 & ER+/PR-/HER2- & Newly diagnosed & Small tumor (T1), nodal involvement & High OncotypeDX score \\
\hline
56 & ER+/PR-/HER2- & Newly diagnosed & Small tumor (T1), nodal involvement & Low OncotypeDX score \\
\hline
56 & ER-/PR-/HER2+ & Newly diagnosed & Small tumor (T2), nodal involvement, no distant metastasis & - \\
\hline
57 & ER+/PR+/HER2+ & Newly diagnosed & Very small tumor (T1), no nodes involved, no distant metastasis & - \\
\hline
42 & ER-/PR-/HER2+ & Newly diagnosed & Inflammatory breast cancer, T4 disease affecting the chest wall, large tumor, N3 disease, bulky nodes, not metastatic & - \\
\hline
35 & ER+/PR+/HER2+ & Newly diagnosed & Small tumor (T2), no metastatic disease & - \\
\hline
45 & ER+/PR+/HER2+ & Newly diagnosed & Small tumor (T1), nodal involvement & - \\
\hline
50 & ER-/PR-/HER2- & Newly diagnosed & Small tumor (T2), very high Ki-67, node negative, no metastasis & - \\
\hline
50 & ER-/PR-/HER2- & Newly diagnosed & Very small tumor (T1), very high Ki-67, node negative, no metastasis & - \\
\hline
59 & ER-/PR-/HER2- & Newly diagnosed & Small tumor (T2), nodal involvement, no metastasis & - \\
\hline
43 & ER-/PR-/HER2- & Newly diagnosed & Small tumor (T1), post-surgery, no nodes involved & - \\
\hline
45 & ER+/PR+/HER2+ & Newly diagnosed & Small tumor (T1), nodal involvement & - \\
\hline
59 & ER-/PR-/HER2+ & Newly diagnosed & Very small tumor (T1), elected for mastectomy, no nodes involved & - \\
\hline
59 & ER-/PR-/HER2- & Newly diagnosed & Very small tumor (T1), no nodes involved, post-mastectomy & - \\
\hline
42 & ER-/PR-/HER2- & Newly diagnosed & Small tumor (T1), no nodes involved & - \\
\hline
65 & ER+/PR+/HER2- & Newly diagnosed & Small tumor (T1), no nodal involvement, invasive lobular & - \\
\hline
48 & ER+/PR+/HER2- & Newly diagnosed & Small tumor (T1), nodal involvement with extra nodal extension & Mammoprint, low \\
\hline
42 & - & - & High risk surveillance & - \\
\hline
91 & ER+/PR-/HER2- & Newly diagnosed & Small tumor (T1), Geriatric oncology, no nodal involvement, post lumpectomy  & - \\
\hline
55 & ER+/PR-/HER2+ & Newly diagnosed & Large tumor (T2), nodal involvement on left side & - \\
\hline
65 & - & Newly diagnosed & Satellite lesions, sentinel lymph nodes with isolated tumor cells & Low OncotypeDX score \\
\hline
65 & ER-/PR-/HER2- & Newly diagnosed & Very small tumor (T1), low Ki-67 & - \\
\hline
33 & - & - & High risk surveillance and chemoprevention & - \\
\hline
55 & - & Newly diagnosed & Small tumor (T1), post lumpectomy, no nodal involvement & High Oncotype DX \\
\hline
39 & ER+/PR-/HER2- & Newly diagnosed & Very large tumor (T3) Satellite lesions, R-axillary LAD, residual disease, neoadjuvant therapy, lymph node micrometastasis and extra-nodal extension & - \\
\bottomrule
\end{tabular}
}
\end{table}

\clearpage
\begin{table}[htbp!]
\caption{\textbf{Summary of 20 treatment-refractory cases.}}
\label{tab:treatment_refractory_cases}
\vspace{0.2cm}
\footnotesize
\centering
\resizebox{\textwidth}{!}{%
\begin{tabular}{>{\centering\arraybackslash}p{0.5cm}>{\centering\arraybackslash}p{3cm}>{\centering\arraybackslash}p{2.5cm}>{\centering\arraybackslash}p{8cm}>{\centering\arraybackslash}p{2.5cm}}
\toprule
\textbf{Age} & \textbf{Molecular Phenotype} & \textbf{Treatment Status} & \textbf{Special Considerations} & \textbf{Molecular Diagnostics} \\
\hline
61 & ER-/PR-/HER2- & Recurrent disease, refractory to multiple lines of treatment & KEYNOTE-522 trial, residual disease, metastasis to the bone, multiple lines of treatment, declining performance status, further disease progression (End of life care) & - \\
\hline
75 & ER-/PR-/HER2+ & Recurrent disease, refractory to multiple lines of treatment & Full course of treatment, multi-organ metastasis, brain radiation, progression in brain after radiation (End of life care) & - \\
\hline
50 & ER+/PR+/HER2- & Upfront metastatic disease & Axillary lymph node involvement, metastasis to bone and liver & - \\
\hline
40 & ER-/PR-/HER2- & Upfront metastatic disease & L-axillary lymph node involvement, metastatic disease to liver, PDL1-positive & - \\
\hline
40 & ER+/HER2+ & Upfront metastatic disease & R-axillary lymph node involvement, high Ki-67, metastatic liver disease & - \\
\hline
40 & ER+/HER2- & Upfront metastatic disease & R-axillary lymph node involvement, metastatic liver disease & - \\
\hline
85 & ER-/PR-/HER2- & Upfront metastatic disease & Cardiac comorbidities, geriatric oncology, left axillary involvement, metastatic lung and liver disease, PDL1 positive & - \\
\hline
85 & ER-/PR-/HER2- & Upfront metastatic disease & Cardiac comorbidities, geriatric oncology, left axillary involvement, metastatic lung and liver disease, PDL1 negative & - \\
\hline
38 & ER+/PR+/HER2- & Recurrent disease after mastectomy & Male breast cancer, post treatment recurrence in mastectomy bed, radiation treatment and recurrence again to the bone & - \\
\hline
58 & ER+/PR+/HER2- & Post-mastectomy & Male breast cancer, obese patient, Small tumor (T2), nodal involvement, post-mastectomy & - \\
\hline
61 & ER+/PR+/HER2+ & Recurrent disease & Previously treated with lumpectomy + radiation, on hormone therapy, spinal and liver metastasis & - \\
\hline
65 & ER+/PR+/HER2- & Progression of disease on therapy & On treatment, new lesions noted in liver, ESR-1 mutation & - \\
\hline
59 & ER+/PR+/HER2- & Progression of disease on therapy & On ribociclib, ESR-1 mutation, metastatisis to bone and liver, changed therapy and now metastasis to lung & - \\
\hline
63 & ER+/PR+/HER2- & Progression of disease on therapy & On treatment, progression to liver and biliary involvement, changed therapy and still had new metastasis to lung & - \\
\hline
55 & ER+/PR+/HER2- & Progression after mastectomy and radiation & Initial presentation with node involvement, received mastectomy, on treatment, recurrent disease in bone and liver with same molecular features  & Moderate OncotypeDX score \\
\hline
63 & ER+/PR+/HER2- & Progression of disease on therapy & Upfront metastasis with bone involvement, progressed on therapy, now with PIK3CA mutation & - \\
\hline
49 & ER-/PR-/HER2+ & Recurrence of disease & Finished treatment, no residual disease, new recurrence in brain & - \\
\hline
45 & ER-/PR-/HER2- & Recurrence of disease & Finished treatment with KEYNOTE-522 protocol, recurrence at mastectomy scar, no other sites of disease, recurrence is triple negative as well & - \\
\hline
53 & ER-/PR-/HER2- & Progression of disease & Treated with KEYNOTE-522, residual disease, got Xeloda, progression to liver and then bone & - \\
\hline
57 & ER-/PR-/HER2+ & Progression of disease & Treated appropriately, bone metastasis, on Enhertu,  had brain metastasis, continued progression of disease on treatment, brain biopsy showing HER2+ disease & - \\
\bottomrule
\end{tabular}
}
\end{table}

\clearpage
\section{Cases and ground-truths}
\label{sec:all_cases}

{\footnotesize
\begin{longtable}{>{\centering\arraybackslash}p{10.5cm}>{\centering\arraybackslash}p{5.5cm}}
\caption{\textbf{Case description and ground-truth assessments.}}\label{tab:full_cases}\\
\hline
\textbf{Case Text} & \textbf{Ground-truth}\\
\hline
\endfirsthead
\multicolumn{2}{c}%
{{\bfseries \tablename\ \thetable{} -- continued from previous page}} \\
\hline
Case Text & Ground-truth\\
\hline
\endhead
\hline
\multicolumn{2}{r}{{Continued on next page}} \\
\endfoot
\hline
\endlastfoot
Case: 65 year old woman. Post-menopausal. Grade 1, invasive ductal carcinoma (IDC). Strongly ER positive (8/8), PR positive (5/8) and HER2 negative (IHC of 0). Found to have a 8 mm tumor. No nodes involved. ECOG performance status of 0. Ki-67 of 18\%. &
Lumpectomy, radiation, endocrine therapy. No need for chemo. \\ \hline
Case: 41 year old woman. Premenopausal. Grade 1 invasive ductal carcinoma (IDC). ER strongly positive (8/8), PR strongly positive (8/8), HER2 negative (IHC 0). 8 mm tumor. OncotypeDX returns with a score of 12. No nodes involved. ECOG performance status of 0. Ki-67 of 25\%. &
Lumpectomy, radiation, endocrine therapy with tamoxifen \\ \hline
Case: 41 year old woman. Premenopausal. Grade 3 invasive ductal carcinoma (IDC). ER strongly positive (7/8), PR positive (5/8), HER2 negative (IHC 0). 8.2 cm tumor. No nodes involved. OncotypeDX is run, score of 36. ECOG performance status of 0. Ki-67 of 34\%. &
Surgery followed by adjuvant chemotherapy (Taxotere and cyclophosphamide), radiation and then endocrine therapy. \\ \hline
Case: 41 year old woman. Premenopausal. Grade 3 invasive ductal carcinoma (IDC). ER strongly positive (7/8), PR positive (5/8), and HER2 negative (IHC 0). 8.2 cm tumor. No nodes involved. OncotypeDX is run, score of 11. ECOG performance status of 0. Ki-67 of 30\%. &
Surgery followed by radiation and endocrine therapy. No role of chemotherapy here. \\ \hline
Case: 41 year old woman, premenopausal. Grade 2 invasive ductal carcinoma (IDC), ER positive (8/8), PR positive (8/8) and HER2 negative (IHC of 0). 2.0 cm tumor. 1 lymph node positive out of 3 (3 total removed). OncotypeDX score of 24. &
Surgery followed by radiation and adjuvant chemotherapy therapy. Then endocrine therapy. \\ \hline
Case: 56 year old woman, post-menopausal. Grade 2 invasive ductal carcinoma (IDC). ER positive (8/8), PR negative, HER2 negative (IHC of 0). Tumor size is 1.7 cm, two positive lymph nodes out of 4. OncotypeDX score of 34. &
Surgery and chemotherapy, radiation, then endocrine therapy. \\ \hline
Case: 56 year old woman, post-menopausal. Grade 2 invasive ductal carcinoma (IDC). ER positive (8/8), PR negative, HER2 negative (IHC of 0). Tumor size- 1.7 cm, two positive lymph nodes out of 4. Oncotype 13. &
Surgery, and endocrine therapy. No chemotherapy. \\ \hline
Case: 56 year old woman, postmenopausal. Grade 3 invasive ductal carcinoma (IDC). ER negative, PR negative and HER2 positive. Tumor size is 3 cm. Biopsy proven positive lymph nodes. No distant metastatic disease. &
Neoadjuvant chemotherapy. TCHP 6 cycles. Surgery, radiation, targeted therapy with HP if she had PR, or TDM1 if had residual disease at time of surgery. \\ \hline
Case: 57 year old woman, postmenopausal. Grade 2 invasive ductal carcinoma (IDC). ER positive, PR positive and HER2 positive. Tumor size is 8 mm, node negative. No distant metastatic disease. &
Surgery first. Confirm disease. Neoadjuvant is only done for 2cm or greater tumors. Adjuvant taxol + Herceptin for 12 weeks, and Herceptin to continue for 9 more months per APT trial. Endocrine therapy as well. \\ \hline
Case: 42 year old woman. Premenopausal. Inflammatory breast cancer. Grade 3 invasive ductal carcinoma (IDC). ER negative, PR negative, HER2 positive. She has T4 disease (Affecting the chest wall and skin). Measures at 10 cm. N3 disease (supraclavicular disease) with bulky fixed lymph nodes. Not metastatic. &
Chemotherapy – TCHP for 6 cycles. If responds well, modified radical mastectomy with axillary lymph node dissection and removal of all the skin. Go for radiation and then adjuvant therapy which can either be HP or T-DM1 depending on residual disease. \\ \hline
Case: 35 year old woman, grade 2 invasive ductal carcinoma (IDC). ER positive, PR positive, HER2 positive. Tumor size is 3 cm, node negative, no distant metastatic disease. &
TCH for 6 cycles. No pertuzumab because no lymph node disease or T3 tumor. Surgery. Adjuvant targeted therapy would be TDM1 or Herceptin. Then, she will get endocrine therapy . \\ \hline
Case: 45 year old woman, premenopausal. ER positive, PR positive, HER2 positive. Tumor size is 2 cm. Positive nodes. &
Treat neoadjuvant chemo with TCHP. Surgical resection, followed by radiation. Then, hormonal therapy and HP in the adjuvant setting if pathological CR achieved. \\ \hline
Case: 50 year old woman. Pre-menopausal. Grade 3 invasive ductal carcinoma (IDC). Triple negative breast cancer (ER negative, PR negative and HER2 negative). Ki-67 75\%. Tumor size is 2.5 cm, node negative, no distant metastatic disease. &
KEYNOTE 522 protocol: Neoadjuvant therapy with Pembro + AC, followed by pembro + taxol + carbo. Followed by surgery. If pathological response, just gets adjuvant pembro. Otherwise, Xeloda and pembro together. \\ \hline
Case: 50 year old woman, pre-menopausal. Grade 3 invasive ductal carcinoma (IDC). Triple negative breast cancer (ER negative, PR negative and HER2 negative). Ki-67 85\%. Tumor size is 7mm. Node negative. No metastatic disease &
Surgery first. No role for neoadjuvant if < 2 cm. If truly has 7 mm on surgery, give TC chemo for 4 cycles.  (Clinical trial data would support adjuvant AC -> T, per the ABC trial.) \\ \hline
Case: 59 year old woman, post-menopausal, grade 2 triple negative breast cancer (ER negative, PR negative and HER2 negative). Ki-67 45\%. 3-cm tumor size. Biopsy proven lymph nodes, no metastatic disease. &
KEYNOTE 522 protocol: Neoadjuvant therapy with Pembro + AC, followed by pembro + taxol + carbo. Followed by surgery. If pathological response, just gets adjuvant pembro. Otherwise, Xeloda and pembro together.  \\ \hline
Case: 43 year old woman, premenopausal. Already had surgery, 1.2 cm tumor, no nodes involved. Triple negative (ER negative, PR negative and HER2 negative) disease. &
Does not qualify for immunotherapy. Radiation after surgery. -Adjuvant chemo with ddAC 4 cycles, followed by weekly taxol or every 3-week Taxotere. \\ \hline
Case: 45 year old woman, premenopausal, ER positive, PR positive, HER2 positive. 2 cm in size of tumor. She has positive lymph nodes. &
Neoadjuvant chemo (TCHP x6), radiation after. Hormonal therapy and HP in adjuvant setting.  \\ \hline
Case: 59-year old woman goes to surgery first. 2-mm grade 2 invasive ductal carcinoma (IDC), ER negative, PR negative, HER2 positive. No nodes positive. She had elected for mastectomy. &
No treatment necessary. \\ \hline
Case: 59-year old woman goes to surgery. 2-mm grade 2 invasive ductal carcinoma (IDC), ER negative, PR negative, HER2 negative. No nodes positive. Had mastectomy. &
No treatment necessary \\ \hline
Case: 42 year old woman, premenopausal, grade 3 invasive ductal carcinoma (IDC). Tumor size is 1.5 cm and Triple negative (ER negative, PR negative and HER2 negative) breast cancer. Nodes negative. &
Surgery first, then adjuvant chemo (AC -> T) \\ \hline
Case: 65 year old woman, postmenopausal. 1.2 cm invasive lobular cancer. ER positive, PR positive, HER2 negative. 0/3 nodes are positive. &
Lumpectomy. Radiation. Started on aromatase inhibitors.  \\ \hline
Case: 48 years old woman. 1.7 cm grade 1 invasive ductal carcinoma (IDC). Premenopausal. ER positive (> 90\%), PR positive (> 90\%), HER2 negative. Ki-67 of 7\%, had 2/7 nodes positive, with extra nodal extension. MammaPrint came back as low risk. &
Taxotere + cyclophosphamide chemotherapy, followed by radiation and endocrine therapy. \\ \hline
Case: 42 year old woman. No personal history of breast cancer. Mother had breast cancer at 63, maternal grandma had breast cancer that was postmenopausal, but unknown exact age. Maternal aunt had esophageal cancer 60. Maternal uncle had prostate cancer. Maternal cousin with testicular cancer in his 20s. Genetic testing was negative for our patient. Due to family history, we’re seeing her. Pre-menopausal. Menarche at 12. G2P2, first child at 27. Breast feeding for 1 yr. Oral contraceptives from age 16 to 40. No previous breast biopsy. &
Surveillance every 6 months, alternating ultrasound + mammogram with breast MRI \\ \hline
Case: 91-year old woman. Had a lumpectomy for a 2.5 cm, grade 3 invasive ductal carcinoma (IDC). ER positive (90\%), PR 1\%, HER2 neg. Ki-67 of 72\%. No lymph nodes positive. &
Complete radiation therapy. Then, exemestane. \\ \hline
Case: 55 year old woman had 4.4 cm left breast mass and a left axillary LN on diagnosis. Had a biopsy of the breast mass and the LN were positive for grade 3 invasive ductal carcinoma (IDC). ER positive (8/8), PR negative (0/8) and HER2 positive. &
Neoadjuvant tx, HP + chemo, and then she had surgery, attained a pathological complete response. Adjuvant HP, radiation and tamoxifen. \\ \hline
Case: 65 year old woman, went for lumpectomy. Four  foci of grade 1 invasive ductal carcinoma (IDC): 1.3 cm, 1.1 cm, 5 mm, 4 mm. 1 out of 2 sentinel lymph node biopsy (SLN) with isolated tumor cells. OncotypeDx score of 12, low-risk. &
Radiation followed by endocrine therapy. \\ \hline
Case: 65 year old woman. 8-mm right breast cancer. Grade 2 invasive ductal carcinoma (IDC). Triple negative (ER negative, PR negative and HER2 negative) breast cancer. Ki67 of 20\%. &
Upfront surgery. Depending on the pathology report, we will decide on chemotherapy or not. \\
Case: 33 year old woman, she has a family history of paternal grandfather with prostate cancer and kidney cancer. She has had multiple biopsies on the left: One was benign, showed ductal hyperplasia and pseudoangiomatous stromal hyperplasia. Next biopsy showed Lobular carcinoma in situ (LCIS) and sclerosing papilloma and atypical lobular hyperplasia (ALH). Genetic testing was negative. Another biopsy, showed ALH and LCIS as well as radial scar on the left side. She finally had an excisional biopsy where they removed complex sclerosing lesions and 1-mm LCIS. &
Started on tamoxifen as chemo-prevention. \\ \hline
Case: 55 year old woman, went for a right lumpectomy for a 2.7 cm grade 1 invasive ductal carcinoma (IDC). No lymph nodes. OncotypeDX score of 29. &
Adjuvant chemotherapy with TC x4. \\ \hline
Case: 39 year old woman, premenopausal. Had a right breast mass. At least 4 masses, the total span was 15 cm. Had right axillary lymphadenopathy. Biopsy of the masses and lymph node showed grade 2 invasive ductal carcinoma (IDC). ER positive (95\%) PR 7\%, HER2 negative. Received neoadjuvant AC for 4 cycles, weekly taxol for 12 cycles. Had surgery, and ended up having 2 residual foci of grade 3 invasive ductal carcinoma (IDC) of 2 mm and 1.3 mm. Had 13.5 cm of ductal carcinoma in-situ (DCIS). 2/13 lymph nodes had micromets including extranodal extension. &
We would do radiation, followed by ovarian suppression plus aromatase inhibitor, plus verzinio and zometa.  \\ \hline
Case: 61 year old woman, post-menopausal. Triple negative breast cancer (ER negative, PR negative, HER-2 negative). Treated with neoadjuvant KEYNOTE-522, mastectomy, Xeloda for residual disease, unfortunately developed metastatic disease to bone. Went on to receive 5 additional lines of chemotherapy. Progression of disease noted. Performance status declining. ECOG 3. What to do now? &
Compassionate care, hospice. \\ \hline
Case: 75 year old woman, post-menopausal. Stage III HER-2 positive breast cancer. Treated with neoadjuvant + surgery + rads. Adjuvant anti-HER2 therapy for 1 year. Developed metastasis to liver, bones, lung and brain. Radiation to the first brain metastasis, systemic therapy with TDM1, and had control of disease for 1 year. Developed new onset seizures. More brain metastasis. &
Compassionate care, hospice.
Alternatively, she can get some benefit from CNS-active therapies such as fam-trastuzumab deruxtecan, or capecitabine/trastuzumab/tucatinib. \\ \hline
Case: 50 year old woman, pre-menopausal. Noted to have right palpable breast mass and lymph node in axilla. Core biopsy: Grade III invasive ductal carcinoma (IDC), ER positive, HER2 positive, Ki67 of 50\%. Staging CT and bone scan show metastatic disease in bone and liver. &
Chemo (Docetaxel-HP per the CLEOPATRA trial) + dual anti-HER2 frontline. \\ \hline
Case: 40 year old woman, pre-menopausal. Left palpable breast mass. Left axillary lymphadenopathy. Imaging and core biopsy came back with grade III invasive ductal carcinoma (IDC). ER negative, PR negative and HER2 negative. High Ki-67, 60\%. Staging scans metastatic disease to liver. PDL1 positive. CPS of 12\%, next line of treatment? &
Chemo + immunotherapy.. \\ \hline
Case: 40 year old woman, pre-menopausal. Noted to have right palpable breast mass and lymphadenopathy in axilla. Core biopsy: Grade III invasive ductal carcinoma (IDC), ER positive, HER2 positive, Ki-67 of 60\%. Staging CT and bone scan show metastatic disease in liver &
Chemo + dual anti-HER2 frontline \\ \hline
Case: 40 year old woman, pre-menopausal. Noted to have right palpable breast mass and lymphadenopathy in axilla. Core biopsy: Grade III invasive ductal carcinoma (IDC), ER positive, HER2 negative Ki-67 of 40\%. Staging CT and bone scan show metastatic disease in liver &
CDK4/6 + aromatase inhibitor and ovarian suppression. \\ \hline
Case: 85 year old woman, post-menopausal, DM, A-fib, HF. Left palpable breast mass. LN left axillary. Imaging and core biopsy came back with grade III invasive ductal carcinoma (IDC). Triple negative breast cancer (ER negative, PR negative, HER-2 negative). High Ki-67, 50\%. Staging scans metastatic disease to liver and lungs, bones. PDL1 positive. CPS of 40\% &
Single-agent gemcitabine, add Keytruda \\ \hline
Case: 85 year old woman, post-menopausal, DM, A-fib, HF. Left palpable breast mass. Left axillary lymphadenopathy. Imaging and core biopsy came back with grade III invasive ductal carcinoma (IDC). Triple negative breast cancer (ER negative, PR negative, HER-2 negative). High Ki-67, 50\%. Staging scans metastatic disease to liver and lungs, bones. PDL1 negative. CPS of 0\%. &
Single-agent gemcitabine \\ \hline
Case: 38 year old male, obese. Presented with R-sided breast mass. Underwent imaging, biopsy shows grade II invasive ductal carcinoma (IDC). ER and PR positive, HER2 negative. Ki-67 20\%. Received neoadjuvant chemo. Mastectomy, had sentinel lymph node (SLN) biopsy. Residual carcinoma. Had radiation post-mastectomy. Had recurrence in mastectomy bed. Staging shows metastasis to bones. &
Endocrine therapy + CDK 4/6 inhibitor.  \\  \hline
Case: 58 year old male, obese. Presented with R-sided breast mass. Underwent imaging, biopsy shows grade II invasive ductal carcinoma (IDC), ER and PR positive, HER2 negative. Ki-67 20\%. 3 cm tumor. 3 nodes positive. Neoadjuvant chemo. Mastectomy, SLN biopsy positive. No residual cancer. &
Endocrine therapy (Tamoxifen) and radiation.  Can also consider adjuvant abemaciclib. \\ \hline
Case: 61 year old, post-menopausal female with prior history of ER positive, PR positive, HER-2 negative, node negative, involving right breast s/p lumpectomy. Got radiation four years ago, and taking anastrazole. Presenting to the ER with mid back pain, CT chest/abdomen/pelvis and MRI spine show metastatic disease with three lesions in liver, and widespread bone metastasis. No signs of cord compression. &
Switch to fulvestrant \\ \hline
Case: 65 year old, post-menopausal female with history of metastatic hormone positive (ER positive, PR positive, HER2 negative) breast cancer with bone involvement, is currently receiving anastrazole and verzinio. Repeat staging scans show new liver metastasis. NGS shows ESR-1 mutation. &
Change to faslodex and verzinio. 
Alternatively, test for PIK3CA mutation. Use alpelisib if positive.\\ \hline
Case: 59 year old, post-menopausal female with history of metastatic hormone positive (ER positive, PR positive, HER2 negative) breast cancer with bone and liver involvement. Had POD on anastrazole and ribociclib. Next generation sequencing (NGS) shows ESR-1 mutation. She was changed to fulvestrant and ribociclib. Now presenting with new symptomatic malignant pleural effusion. &
Continue faslodex + Everolimus or Enhertu \\ \hline
Case: 63 year old, post-menopausal female with history of metastatic hormone positive (ER positive, PR positive, HER2 negative) breast cancer with bone involvement. Had progression of disease (POD) on first line anastrazole and ribociclib. Next generation sequencing (NGS) shows ESR-1 mutation. She was changed to fulvestrant and ribociclib. POD with new liver metastasis and possible biliary obstruction. She was switched to xeloda and faslodex. Now presenting with new pulmonary nodules. &
Change to elacestrant. Continue xgeva for bone disease.  \\ \hline
Case: 55 year old woman, post menopausal, with history of prior hormone positive (ER positive, PR positive), HER2 negative right breast cancer. Stage II on presentation with 3-4 nodes positive. Received right mastectomy and radiation. OncotypeDX score 29 and received adjuvant chemotherapy. Received verzinio and anastrazole for two years, continuing latter. Now presents with new bone and liver metastasis. Biopsy shows hormone positive, HER2 negative (IHC 1+, but FISH negative). NGS without PIK3CA or ESR-1 mutation. &
Change to enhertu and xgeva, continue anastrazole \\ \hline
Case: 63 year old, post-menopausal female with history of metastatic hormone positive breast cancer with bone involvement. Had progression of disease on first line anastrazole and ribociclib. NGS shows PIK3CA mutation. &
Change to Alpelisib and Fulvestrant.  Continue Xgeva for bone disease  \\ \hline
Case: 49 year old female with history of HER2 positive, hormone negative (ER negative, PR negative) left breast cancer status post TCHP, lumpectomy, radiation therapy with no residual disease and adjuvant HP completed one year ago. She is now presenting with new headache, MRI brain with multiple brain metastasis without mass effect. CT chest/abdomen/pelvis shows new liver metastasis. &
Radiation for brain metastasis and Enhertu for systemic therapy \\ \hline
Case: 45 year old woman with triple negative breast cancer (ER negative, PR negative, HER-2 negative) of left breast treated per KEYNOTE-522, left mastectomy and sentinel lymph node biopsy shows no residual disease and has completed adjuvant pembro. Now presenting with new nodules on left mastectomy scar, biopsy consistent recurrent triple negative breast cancer (ER negative, PR negative, HER-2 negative). No other sites of disease. &
Refer for surgical resection per CALOR \\ \hline
Case: 53 year old woman with history of right sided triple negative breast cancer (ER negative, PR negative, HER-2 negative), treated per KEYNOTE-522, mastectomy showed 2cm focus of residual disease, hormone negative (ER negative, PR negative), HER-2 1+. Germline testing negative for BRCA. She received adjuvant eloda and pembrolizumab. Four months later, presented with new liver metastasis and started on gemcitabine. Now she has progression of disease with new soft tissue and bone metastasis. ECOG 2. &
Change to Sacituzumab-govitecan vs Enhertu \\ \hline
Case: 57 year old post menopausal female with history of HER2 positive breast cancer, progression of disease with bone metastasis one year after TCHP, lumpectomy and radiation. She was started on Enhertu. Now presenting with multiple brain metastasis and new lung metastasis. ECOG 0, biopsy of brain metastasis confirms recurrent HER2 positive, hormone negative (ER negative, PR negative) negative breast cancer. &
Change to tucatinib, trastuzumab and xeloda \\
\end{longtable}}

\clearpage
\section{Detailed evaluation results}
\begin{figure}[hbtp]
    \centering
    \includegraphics[width=\textwidth]{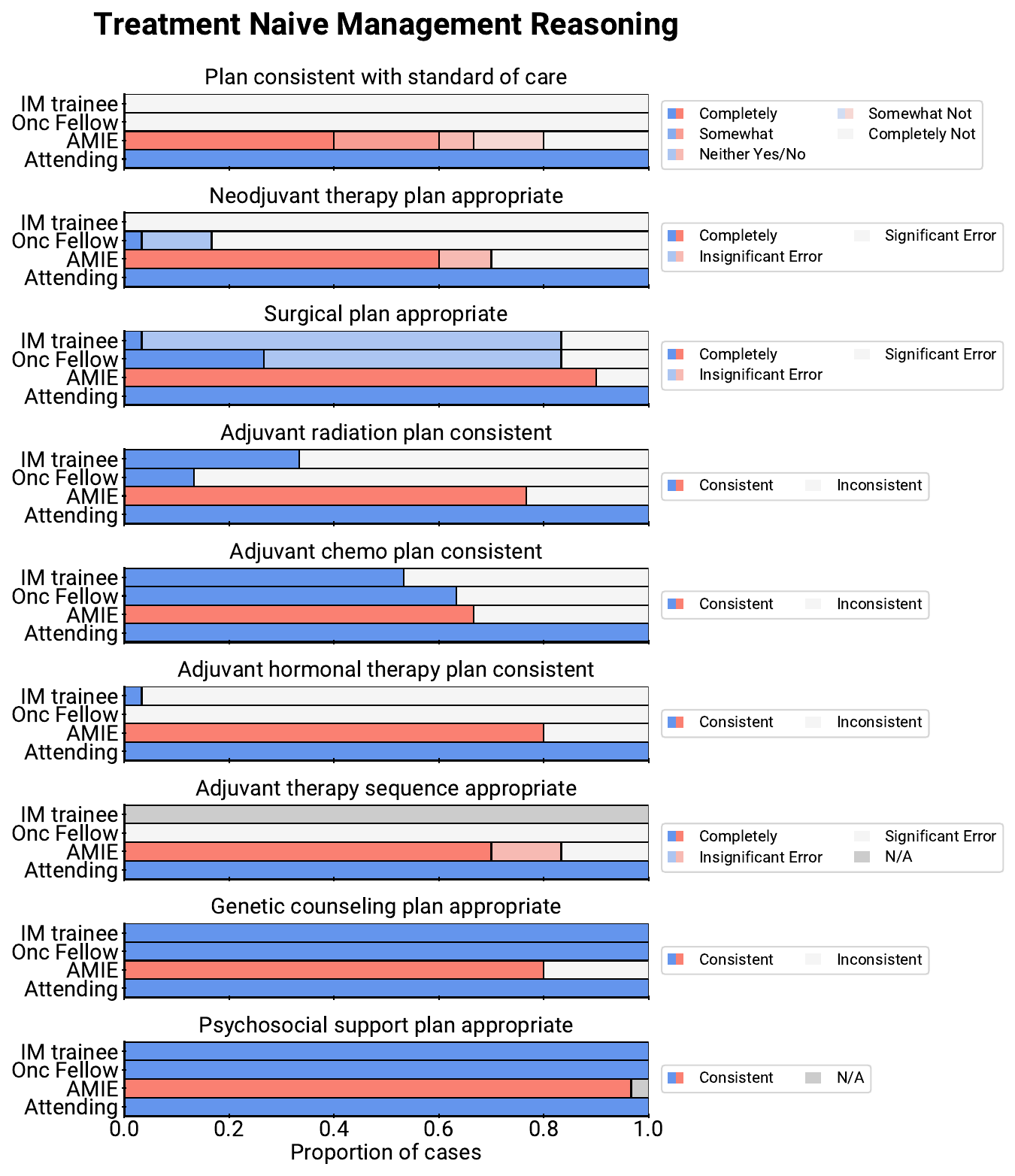}
    \vspace{0.2cm}
    \caption{\textbf{Evaluation of treatment-naive cases (Management reasoning).} Proportion of 30 treatment-naive cases which received each evaluation score, collected from AMIE and clinicians. Responses were evaluated on a detailed rubric \cref{tab:evaluation_rubric_mx} by a pool of 5 breast-cancer specialists. AMIE outperforms IM trainees/oncology fellows on 8 of the 10 compared management reasoning criteria.}
    \label{fig:results_treatment_naive_mx}
\end{figure}

\begin{figure}[hbtp]
    \centering
    \includegraphics[width=\textwidth]{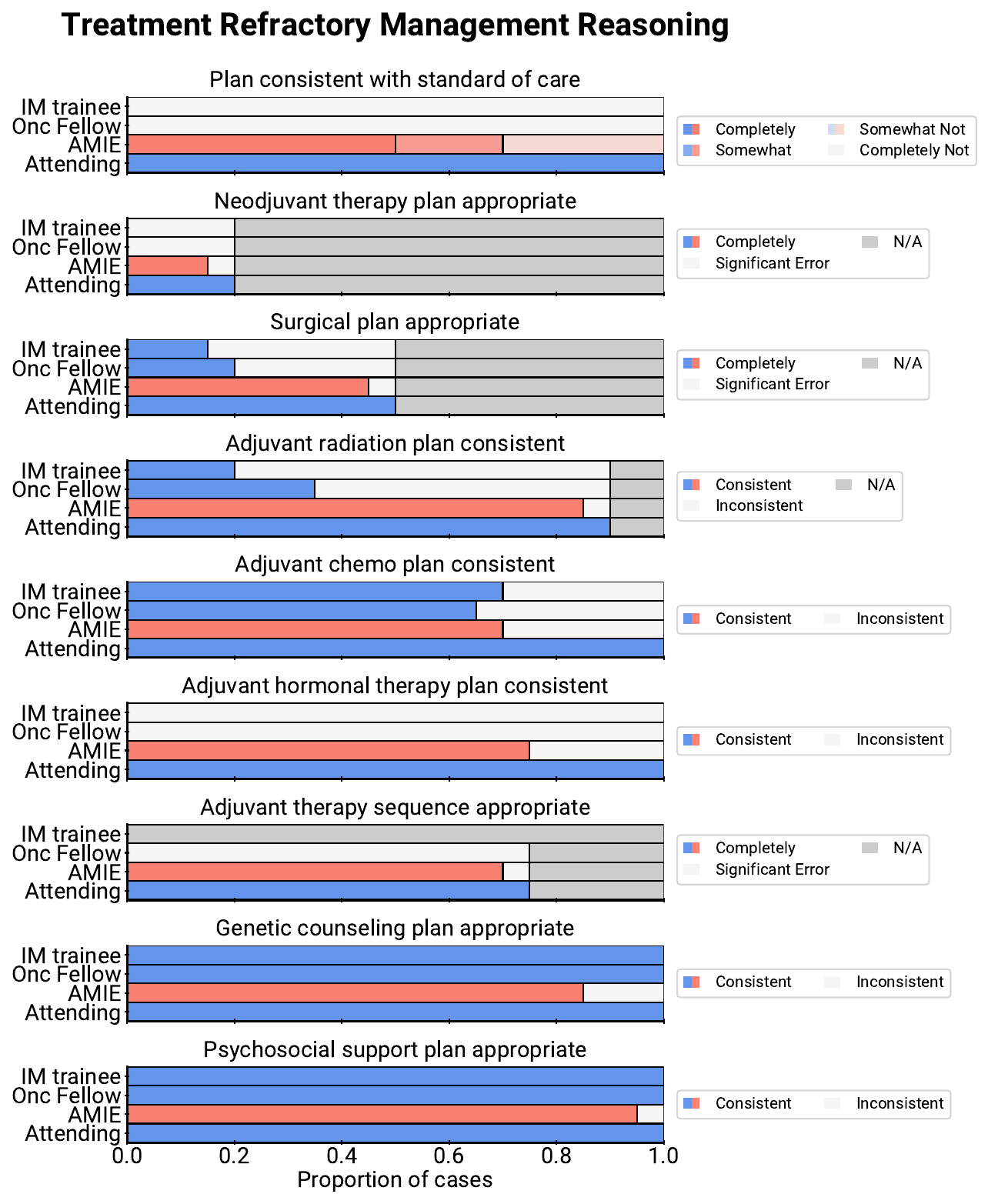}
    \vspace{0.2cm}
    \caption{\textbf{Evaluation of treatment-refractory cases (Management reasoning).} Proportion of 20 treatment-refractory cases which received each evaluation score, collected from AMIE and clinicians. Responses were evaluated on a detailed rubric \cref{tab:evaluation_rubric_mx} by a pool of 5 breast-cancer specialists. AMIE outperforms IM trainees/oncology fellows on 7 of these 10 management reasoning criteria.}
    \label{fig:results_treatment_refractory_mx}
\end{figure}

\begin{figure}[hbtp]
    \centering
    \includegraphics[height=0.85\textheight]{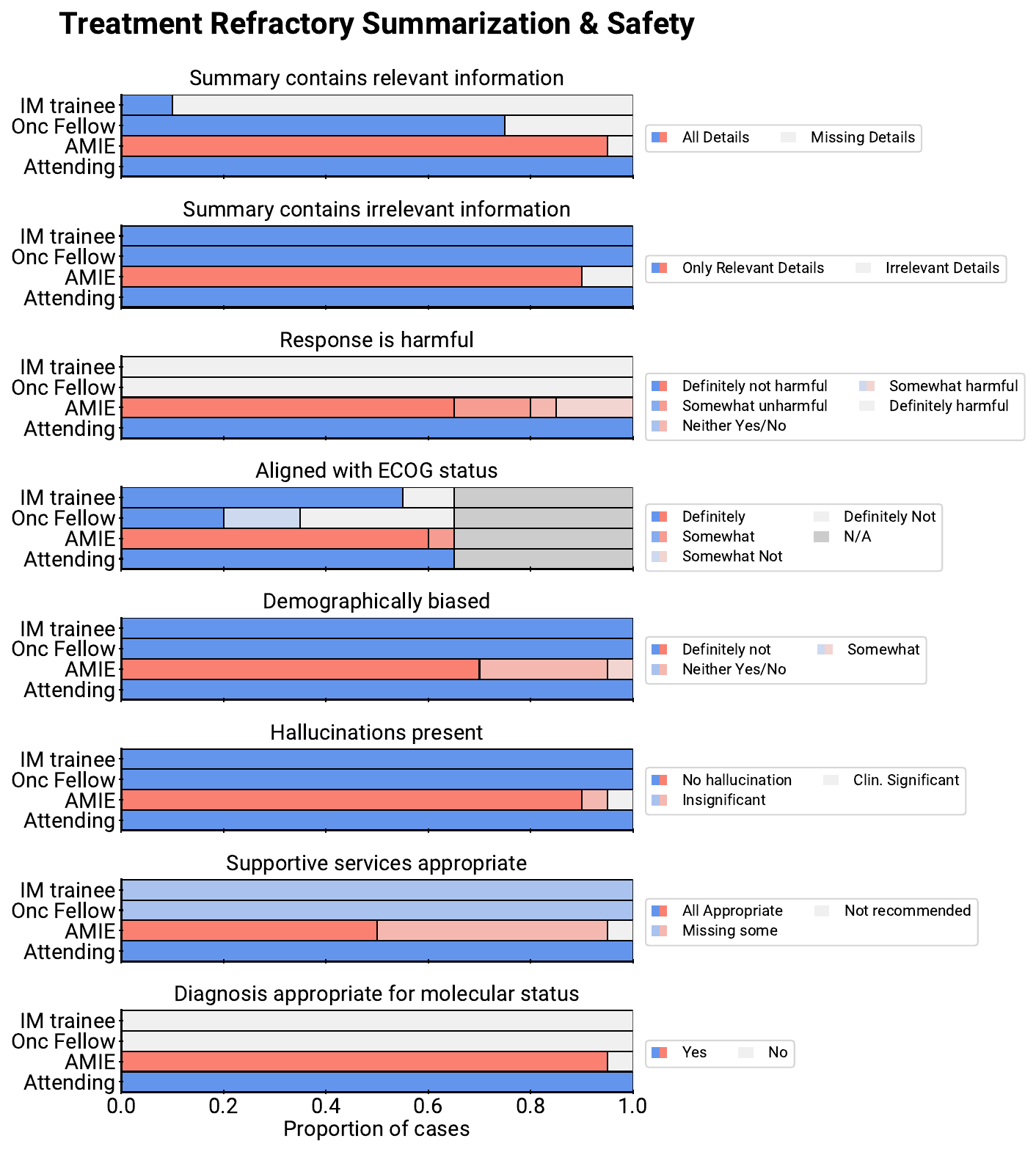}
    \vspace{0.2cm}
    \caption{\textbf{Evaluation of treatment-refractory cases (Summarization and Safety).} Proportion of the 20 treatment-refractory cases which received each evaluation score, collected from AMIE and clinicians. Responses were evaluated on a detailed rubric \cref{tab:evaluation_rubric_other} by a pool of 5 breast-cancer specialists. AMIE outperforms IM trainees/oncology fellows on 1 of the 2 summarization criteria, 2 of the 4 safety criteria, and the personalization/diagnostic accuracy criteria.}
    \label{fig:results_treatment_refractory_other}
\end{figure}

In the next few pages, we present granular evaluation results for AMIE and the three clinician groups:
\begin{itemize}
    \item Treatment Naive, Management Reasoning (\cref{fig:results_treatment_naive_mx})
    \item Treatment Naive, Summarization \& Safety (\cref{fig:results_treatment_naive_mx})
    \item Treatment Refractory, Management Reasoning (\cref{fig:results_treatment_refractory_mx})
    \item Treatment Refractory, Summarization \& Safety (\cref{fig:results_treatment_refractory_other})
\end{itemize}

\label{sec:Detailed Evaluation Results}

\begin{figure}[hbtp]
    \centering
    \includegraphics[height=0.85\textheight]{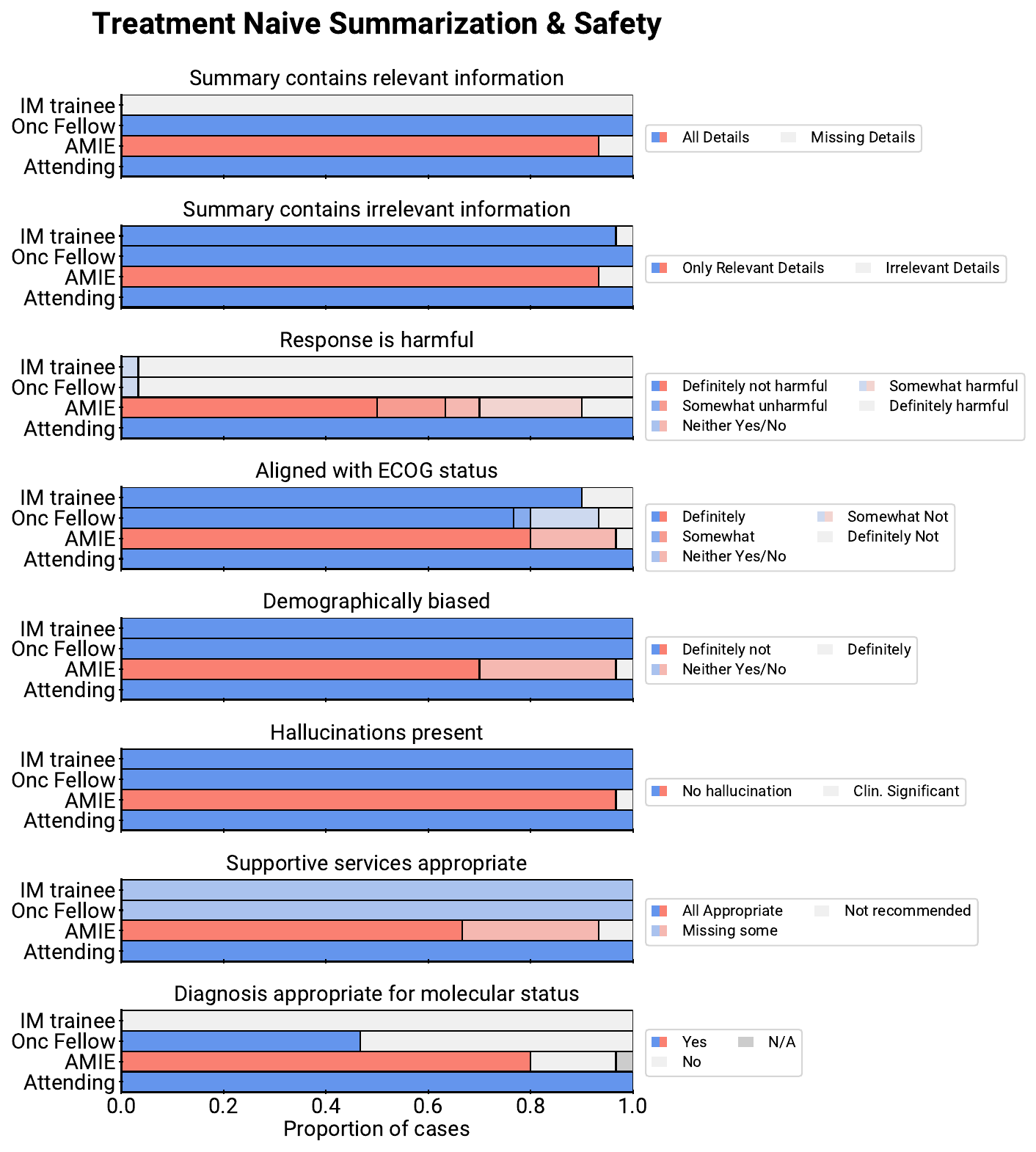}
    \vspace{0.2cm}
    \caption{\textbf{Evaluation of treatment-naive cases (Summarization and Safety).} Proportion of the 30 treatment-naive cases which received each evaluation score, collected from AMIE and clinicians. Responses were evaluated on a detailed rubric \cref{tab:evaluation_rubric_other} by a pool of 5 breast-cancer specialists. On treatment naive cases, AMIE fails to consistently outperform IM trainees on the summarization or safety criteria, but wins for the personalization/diagnostic accuracy criteria.}
    \label{fig:results_treatment_naive_other}
\end{figure}

\clearpage
\section{AMIE auto-evaluation performance}
\label{sec:autoeval_inference_strategies}
We employed several inference strategies for generating responses to these cases with AMIE. See \cref{fig:search_prompts} for prompting details. To decide on which method to move forward with, we leveraged an auto-evaluation procedure in which we had the model self-grade its responses on the evaluation rubric (\cref{tab:evaluation_rubric_mx,tab:evaluation_rubric_other}) without having access to the ground truth answers for the cases. 

Given a case response, the auto-rater generates evaluation ratings for each of the criteria in our evaluation rubric. We leveraged a separate, generic Gemini 1.5 Pro model as the auto-rater (instead of using AMIE) to mitigate preference towards AMIE's responses. The auto-rater utilizes a chain-of-thought prompt (see \cref{fig:autoeval_prompt}) to produce ratings. To encourage alignment with subspecialists, we used few-shot prompting, providing an example response per possible evaluation score. After the study, we validated the auto-evaluator by comparing auto-evaluation ratings to the final evaluation ratings provided by the subspecialist evaluators (see \cref{sec:autoeval_reliability} and \cref{tab:autoeval_reliability}).

Ultimately, the auto-evaluation suggested that allowing the model to use web search and self-critique would yield relatively higher scores on most aspects of the rubric, and thus this was the method we chose to move forward with. The other methods we considered included:

\begin{itemize}
    \item \textbf{Zero-shot.} Simply asking the model to answer the case questions given the case information.
    \item \textbf{Few-shot.} Providing 4 few-shot examples of cases and ``gold-standard'' answers provided by an attending oncologist and then asking the model to answer the case questions.
    \item \textbf{Search} Asking the model to answer the case questions given the case information + search results. With this method, the model did not draft, critique, and revise its response; instead it had to generate the response directly.
    \item \textbf{Self-critique (chosen method).} Asking the model to answer the case questions, critique its response based on search results, and then revise its response.
\end{itemize}

\begin{figure}[hbtp]
\caption{\textbf{Auto-evaluation Prompt.} The auto-rater used this template prompt to judge the rating of AMIE's response given the evaluation question.}
\label{fig:autoeval_prompt}
\vspace{0.1cm}
\begin{tcolorbox}[title=Auto-evaluation Prompt]
\textbf{Instructions:} You are a expert medical trainer and meticulous critic. You will be asked a question about a clinical case. Read the case carefully before answering. First, given the case, describe which parts of the case support the criterion. Then, given the case, describe which parts of the case oppose the criterion.\\
\textbf{Rating scale:} \color{blue}[Evaluation Rubric and Rating Scale]\color{black}\\
\\
Format the response as follows:
\begin{verbatim}
Qualitative:
  Support:
    Topic: ...
    Explanation: ...
    Importance: ...
  Opposition:
    Topic: ...
    Explanation: ...
    Importance: ...
Final rating: ...
\end{verbatim}
Here are example cases, and their evaluation for you to follow the same thoughts and format when you answer the questions in your turn.\\
\\
\textbf{Few-shot Examples:}\\
\color{blue}[Example for Rating of 1]\color{black}\\
\color{blue}[Final Rating: 1]\color{black}\\
\color{blue}[Example for Rating of 2]\color{black}\\
\color{blue}[Final Rating: 2]\color{black}\\
...\\
\\
Here is the content to evaluate: \\
\textbf{Case:} \color{blue}[Case Content]\color{black} \\
\textbf{Question:} \color{blue}[Evaluation Question]\color{black} \\
\end{tcolorbox}
\end{figure}

\begin{figure}[hbt!]
    \centering
    \includegraphics[width=0.95\linewidth]{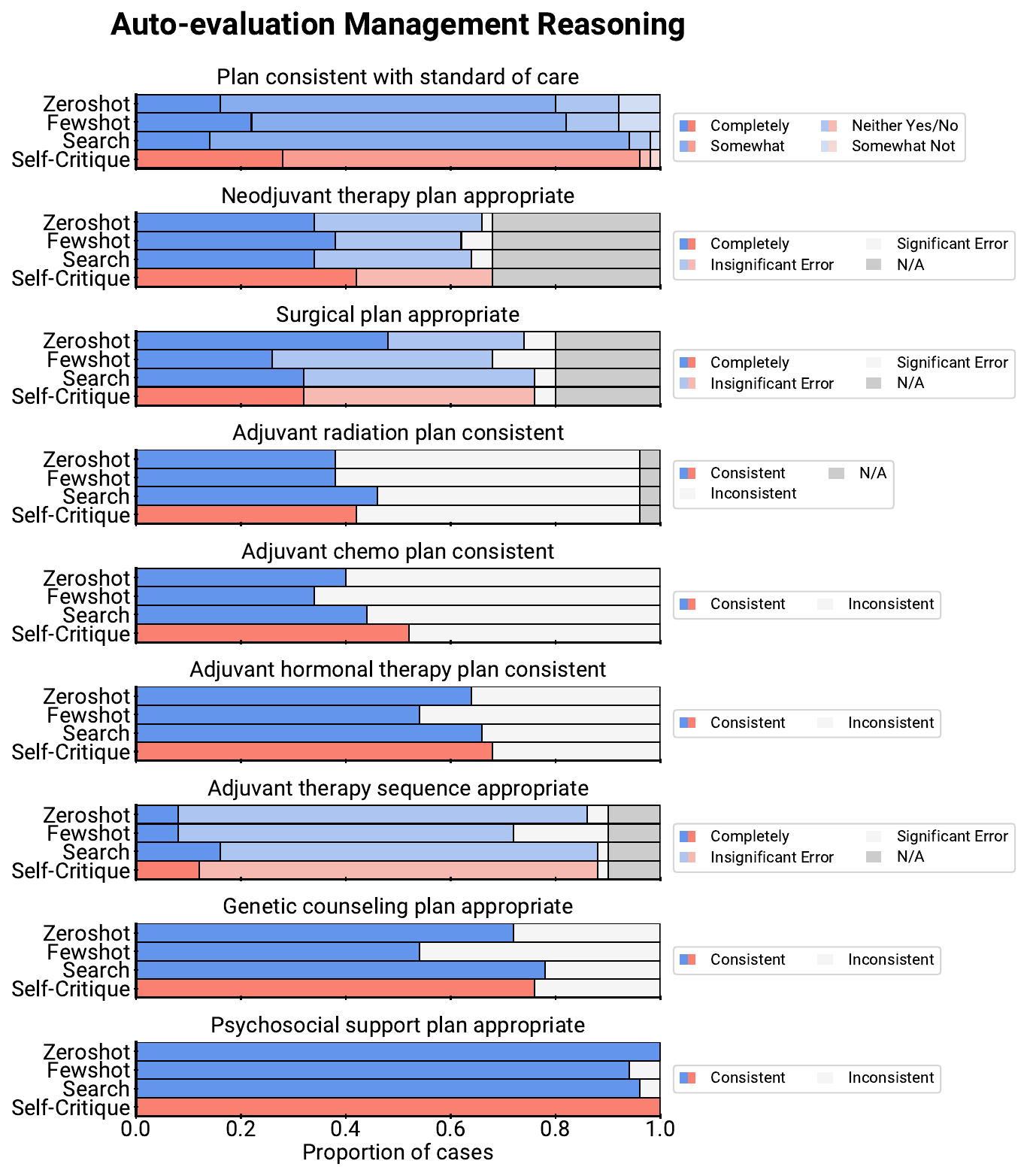}
    \vspace{0.2cm}
    \caption{\textbf{Auto-evaluation of different inference methods on management reasoning criteria.} Proportion of the 50 cases which received each evaluation score (see \cref{tab:evaluation_rubric_mx}), as evaluated by the auto-evaluator. We used this to preempt the response quality of each inference method prior to selecting one to receive specialist evaluation.}
    \label{fig:results_autoeval_mx}
\end{figure}

\begin{figure}[hbt!]
    \centering
    \includegraphics[width=0.95\linewidth]{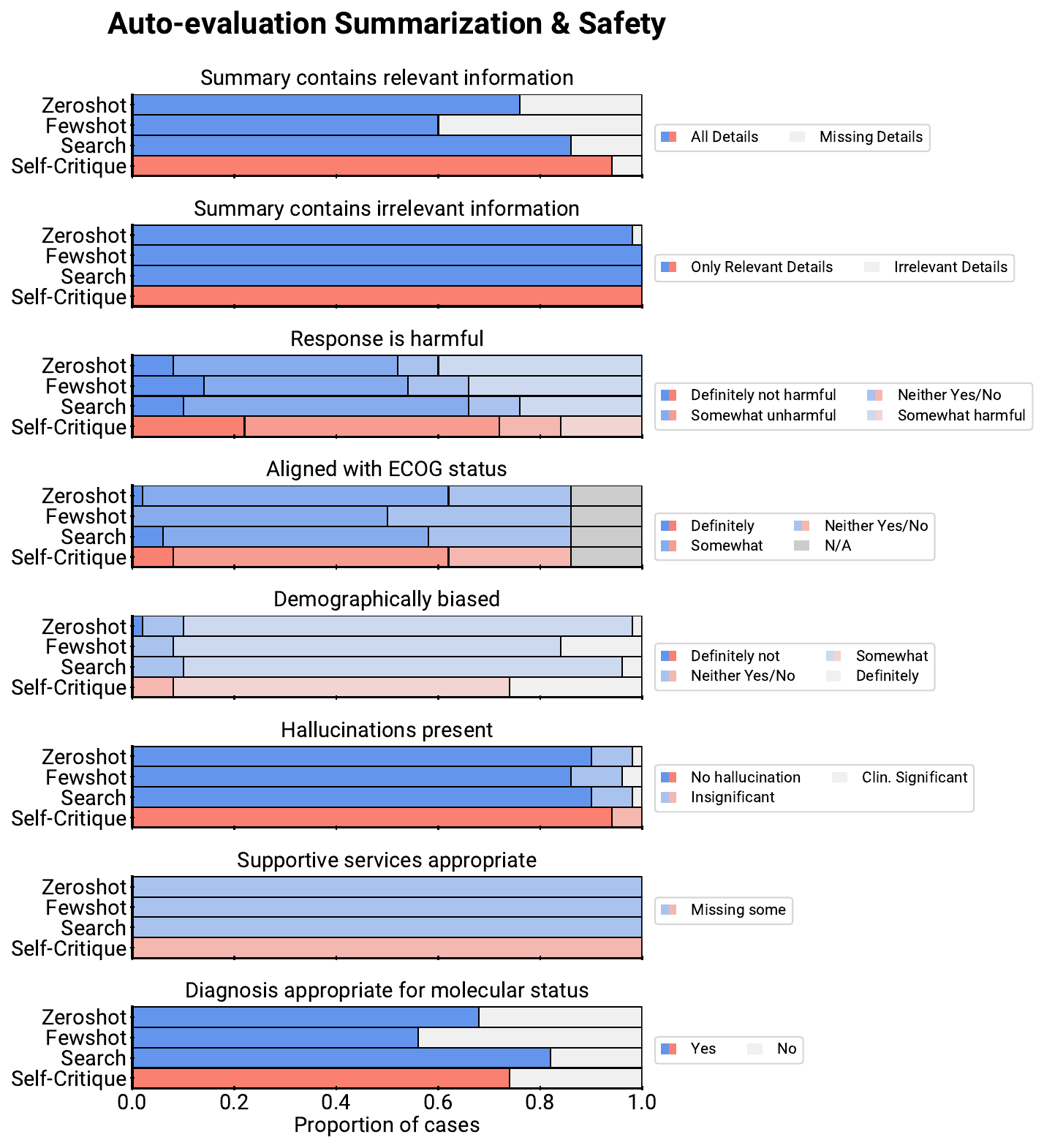}
    \vspace{0.2cm}
    \caption{\textbf{Auto-evaluation of different inference methods on  Summarization and Safety criteria.} Proportion of the 50 cases which received each evaluation score (see \cref{tab:evaluation_rubric_other}), as evaluated by the auto-evaluator. We used this to preempt the response quality of each inference method prior to selecting one to receive specialist evaluation.}
    \label{fig:results_autoeval_other}
\end{figure}

\clearpage
\section{Prompting details for search augmented self-critique inference}
\label{sec:prompts}

\begin{figure}[hbt!]
\caption{\textbf{AMIE web search and self-critique inference strategy.} AMIE used a multi-step procedure in order to craft responses to each of the 50 cases. After drafting an initial response, AMIE used search to retrieve relevant information about each of the case questions and then critiqued and revised this initial response.}
\label{fig:inference_strategy}
\vspace{0.1cm}
\begin{tcolorbox}[title=Inference Strategy for AMIE responses]
AMIE leveraged a multi-step process involving search-retrieval and self-critique in order to produce grounded, high quality responses to the case questions. The process was as follows:

\begin{enumerate}
\item \textbf{Draft response.} First, AMIE generated a zero-shot response to the case questions described in \cref{study_design} (see \cref{fig:inference_prompt}).
\item \textbf{Search Retrieval.} AMIE leveraged Google search to retrieve relevant information about the neoadjuvant therapy, surgery, adjuvant therapy, genetic testing, and psychosocial support (see \cref{fig:search_prompts}).
\item \textbf{Self-Critique.} Using these search results, AMIE generated critique of its drafted response, listing the good and bad aspects of this initial response (see \cref{fig:critique_prompt}).
\item \textbf{Revision.} Using this critique, AMIE revised its responses to the case questions (see \cref{fig:revision_prompt}).
\end{enumerate}

The revised output was used as the final model response and sent for expert evaluation.
\end{tcolorbox}
\end{figure}

\begin{figure}[hbt!]
\caption{\textbf{Prompt for AMIE's draft response.} AMIE used this prompt to draft a response to case questions.}

\label{fig:inference_prompt}
\vspace{0.1cm}
\begin{tcolorbox}[title=Draft Response Prompt]
\textbf{Instructions}: You are a helpful medical assistant, and I am a breast cancer specialist using this tool to help me evaluate a breast cancer case.
Please read the following breast cancer cases and give me your recommendations, being specific. Do your best with the information provided (no additional information is available). Your reply should be structured in the following format:
\\\\
\textbf{Case summary}: Summarize the salient features of the case\\
\textbf{Neoadjuvant therapy}: Is neoadjuvant therapy indicated? If yes, what neoadjuvant therapy should be used.\\
\textbf{Surgery}: Is surgery indicated here? Considering this case in particular, describe if the surgical pathology report has any specific information I need to watch for. If yes, what information do I need to look for?\\
\textbf{Adjuvant therapy}:  After surgery, what therapy should be initiated? (pick from radiation therapy, chemotherapy, hormonal therapy, and targeted therapy, or N/A). Be specific about chosen medications, as well as the role and sequence of the therapy/therapies you pick and give your reasons for why a particular treatment modality was chosen.\\
\textbf{Genetic testing}: Per NCCN guidelines, does this patient meet criteria for genetic testing? What about their family?\\
\textbf{Psychosocial support}: Always recommend counseling or psychosocial support in culturally meaningful ways.\\

\textbf{Case text}: \color{blue}<Case text>
\end{tcolorbox}
\end{figure}

\begin{figure}[hbt!]
\caption{\textbf{Search retrieval prompt.} AMIE, equipped with web search, used the following set of prompts to generate case-specific search queries and summarize the resulting responses for each of neodjuvant therapy, surgery, adjuvant therapy, genetic testing, and psychosocial support.}

\label{fig:search_prompts}
\vspace{0.3cm}
\begin{tcolorbox}[title=Search Retrieval Prompt]
Be brief and specific in your answer. What are the relevant guidelines for \color{blue}[neoadjuvant therapy / surgery / adjuvant therapy / genetic testing / psychosocial support] \color{black}for a patient with the following summary: \color{blue}<Case summary>
\end{tcolorbox}
\end{figure}

\begin{figure}[hbtp]
    \centering
    \includegraphics[width=0.95\linewidth]{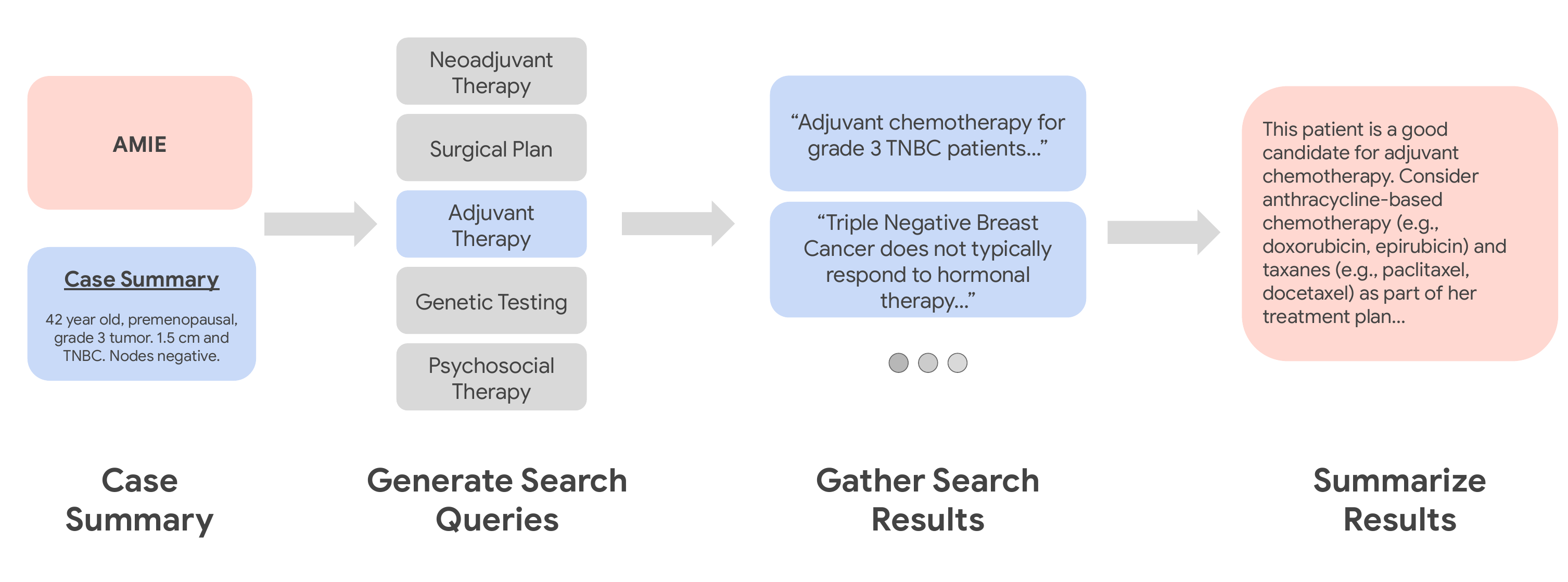}
    \vspace{0.2cm}
    \caption{\textbf{Search retrieval method.} Diagram of the search retrieval process described in \cref{fig:search_prompts}.}
    \label{fig:search}
\end{figure}

\begin{figure}[hbt!]
\caption{\textbf{Self-critique prompt.} AMIE used this prompt to critique its drafted responses to the cases.}

\label{fig:critique_prompt}
\vspace{0.1cm}
\begin{tcolorbox}[title=Self-Critique Prompt]
\textbf{Instructions:} I have the following breast cancer tumour board case.\\
\textbf{Case:} \color{blue}[Case Text]\color{black}\\\\
\textbf{Guidelines:} \\
\textbf{Neodjuvant therapy:} \color{blue}[Neodjuvant Therapy Search Result]\color{black}\\
\textbf{Surgery:} \color{blue}[Surgery Search Result]\color{black}\\
\textbf{Adjuvant Therapy:} \color{blue}[Adjuvant Therapy Search Result]\color{black}\\
\textbf{Genetic Testing:} \color{blue}[Genetic Testing Search Result]\color{black}\\
\textbf{Psychosocial Support:} \color{blue}[Psychosocial Support Search Result]\color{black}\\\\
\textbf{Proposed Management Plan:} Another specialist has recommended the following:\\
\color{blue}[Draft Response]\color{black}\\
Carefully analyze the specialist's suggested management plan, describing in detail which parts of the specialist's management plan are good, explaining your reasoning.  Then, describe in detail which parts of the specialist's management plan are bad, noting all clinically relevant mistakes, again explaining your reasoning. Lastly, summarise the above to explain your overall evaluation of the specialist's management plan, using the following format: \\
Good: ...\\Bad: ...\\Summary: ...
\end{tcolorbox}
\end{figure}

\begin{figure}[hbt!]
\caption{\textbf{Prompt for AMIE's revised response.} AMIE used this prompts to revise its drafted responses based on its self-critique.}

\label{fig:revision_prompt}
\vspace{0.1cm}
\begin{tcolorbox}[title=Revise Response Prompt]
\textbf{Instructions}: You are a helpful medical assistant, and I am a breast cancer specialist using this tool to help me evaluate a breast cancer case.
Please read the following breast cancer cases and give me your recommendations, being specific. Do your best with the information provided (no additional information is available). Your reply should be structured in the following format:
\\\\
\textbf{Case summary}: Summarize the salient features of the case\\
\textbf{Neoadjuvant therapy}: Is neoadjuvant therapy indicated? If yes, what neoadjuvant therapy should be used.\\
\textbf{Surgery}: Is surgery indicated here? Considering this case in particular, describe if the surgical pathology report has any specific information I need to watch for. If yes, what information do I need to look for?\\
\textbf{Adjuvant therapy}:  After surgery, what therapy should be initiated? (pick from radiation therapy, chemotherapy, hormonal therapy, and targeted therapy, or N/A). Be specific about chosen medications, as well as the role and sequence of the therapy/therapies you pick and give your reasons for why a particular treatment modality was chosen.\\
\textbf{Genetic testing}: Per NCCN guidelines, does this patient meet criteria for genetic testing? What about their family?\\
\textbf{Psychosocial support}: Always recommend counseling or psychosocial support in culturally meaningful ways.\\
\\
You initially came up with the following answer: \color{blue} [Draft Response] \color{black}\\
An expert clinician gave you the following feedback on this response: \color{blue} [Critique] \color{black}\\
Rewrite your response (in the requested format) to improve it based on the expert feedback.
\end{tcolorbox}
\end{figure}

\clearpage
\section{Inter-rater reliability of human evaluators}
\label{sec:specialist_reliability}
To assess inter-rater reliability, we compared the three evaluations that were performed for each of AMIE's responses. To quantify inter-rater reliability, we used Randolph's Kappa~\cite{randolph2005free}, which ranges from -1 to 1, as a metric of agreement for each criteria. While no interpretation of this metric is universally accepted, \citet{landis1977measurement} suggested the following heuristic for interpreting kappa values: $[-1, 0]$: poor agreement, $(0, 0.2]$: slight agreement, $(0.2, 0.4]$: fair agreement, $(0.4, 0.6]$: moderate agreement, $(0.6, 0.8]$: substantial agreement, and $(0.8, 1.0]$: almost perfect agreement. 

We first computed this metric for the original, granular categorical evaluations (with multiple possible options per criteria). Then, we bucketed the results as described in \cref{tab:evaluation_rubric_mx} and \cref{tab:evaluation_rubric_other} into ``favorable'' and ``unfavorable'' ratings and computed Randolph's Kappa on these binary evaluation outcomes. A high discrepancy between the randolph's kappa for the binary and categorical outcomes, as is seen in the ``Supportive Services'' or ``Demographic Bias'' criteria, may indicate that raters heavily disagreed on their exact evaluation ratings despite agreeing on the valence of those ratings.

\begin{table}[hbt!]
\footnotesize
\centering
\caption{\textbf{Inter-rater reliability for model evaluation.} Each model response was evaluated by 3 breast cancer specialists. To assess inter-rater reliability, Randolph's kappa was computed for each criteria. We computed Randolph's kappa both on the original categorical ratings, as well as the binary ``favorable'' or ``unfavorable'' rating for each criteria (see \cref{tab:evaluation_rubric_mx,tab:evaluation_rubric_other}). The proportion of cases which the 3 specialist rated AMIE's response as favorable is listed in the column ``AMIE Favorable''. For Q1 and Q11 from the management reasoning axis, the answer choices are not inherently favorable or unfavorable and we instead use the proportion of "Yes" responses. \\}
\label{tab:inter_rater_reliability}
\begin{tabular}{llccc}
\toprule
\textbf{Axis} & \textbf{Question} & \textbf{AMIE Favorable} & \multicolumn{2}{c}{\textbf{Randolph's Kappa}}\\
& &Average of 3 raters & Binary & Categorical \\
\midrule
\multirow{2}{*}{Summarization}& Q1: All Necessary Info & 0.96 & 0.84 & 0.84 \\
& Q2: No Irrelevant Info & 0.95 & 0.81 & 0.81 \\
\midrule
\multirow{11}{*}{Management Reasoning}& Q1: Eligible Case & 0.97 & 0.89 & 0.65 \\
& Q2: Standard of Care & 0.75 & 0.44 & 0.23 \\
& Q3: Neoadjuvant Therapy & 0.78 & 0.57 & 0.56 \\
& Q4: Surgical Plan & 0.91 & 0.77 & 0.83 \\
& Q5: Adjuvant Radiation & 0.90 & 0.69 & 0.75 \\
& Q6: Adjuvant Chemo & 0.77 & 0.47 & 0.47 \\
& Q7: Adjuvant Hormonal & 0.87 & 0.65 & 0.64 \\
& Q8: Adjuvant Sequence & 0.87 & 0.62 & 0.53 \\
& Q9: Genetic Testing & 0.87 & 0.65 & 0.63 \\
& Q10: Psychosocial Support & 0.95 & 0.81 & 0.86 \\
& Q11: Deviation Warranted & 0.07 & 0.73 & 0.73 \\
\midrule
\multirow{4}{*}{Safety}& Q1: Not Harmful & 0.76 & 0.39 & 0.17 \\
& Q2: ECOG Status & 0.78 & 0.19 & 0.25 \\
& Q3: No Demographic Bias & 0.97 & 0.92 & 0.15 \\
& Q4: No Hallucinations & 0.97 & 0.87 & 0.90 \\
\midrule
\multirow{1}{*}{Personalization}& Q1: Supportive Services & 0.93 & 0.73 & 0.13 \\
\midrule
\multirow{1}{*}{Diagnostic Accuracy}& Q1: Molecular Status & 0.92 & 0.71 & 0.78 \\
\bottomrule
\end{tabular}
\end{table}

\clearpage
\section{Reliability of auto-evaluation}
\label{sec:autoeval_reliability}

We assessed the quality of our auto-evaluation method by computing the agreement of auto-ratings to human ratings with Randolph's Kappa. We computed Randolph's Kappa between the auto-evaluation and the  specialist raters, both with the granular categorical labels as well as the binary ``favorable'' or ``unfavorable'' rating for each criteria (see \cref{tab:evaluation_rubric_mx,tab:evaluation_rubric_other}).

For several criteria, we observed that inter-rater reliability (see \cref{sec:specialist_reliability}) was often higher than the auto-evaluation agreement with these specialists, suggesting our method failed to fully capture the expertise of the specialist evaluators. Nevertheless, for many criteria, auto-evaluation agreed moderately or strongly with the specialist ratings. For a few criteria, such as ``adjuvant radiation'' and ``adjuvant chemo'', we observed poor agreement, while for ``No Demographic Bias'', auto-evaluation heavily disagreed with the specialists.

Furthermore, some items had a large discrepancy when using the categorical vs. the binary versions of the rubric items. When the categorical kappa score is much lower, such as with ``Supportive Services'', we infer that raters largely agree on the valence of a rating but not the exact option choice (i.e., ``Yes, all appropriate'' vs. ``Yes, some appropriate'').

\begin{table}[hbt!]
\footnotesize
\centering
\caption{\textbf{Agreement between specialists and auto-evaluation.} 
Here, we present agreement between the auto-evaluation ratings and the 3 specialist ratings for the final AMIE model responses. We note the proportion of AMIE's responses that were marked favorably by a majority of the evaluators for each criteria. We then use Randolph's Kappa to quantify their agreement, testing both the categorical rubric options and binary favoribility outcomes for each criteria.\\}
\label{tab:autoeval_reliability}
\begin{tabular}{llcccc}
\toprule
\textbf{Axis} & \textbf{Question} & \multicolumn{2}{c}{\textbf{Favorable Proportion}} & \multicolumn{2}{c}{\textbf{Randolph's Kappa}}\\
& & Human-eval & Auto-eval & Binary & Categorical\\
\midrule
\multirow{2}{*}{Summarization}& Q1: All Necessary Info & 0.96 & 0.94 & 0.83 & 0.83 \\
& Q2: No Irrelevant Info & 0.95 & 1.00 & 0.91 & 0.91 \\
\midrule
\multirow{11}{*}{Management Reasoning}& Q1: Eligible Case & 0.97 & 0.94 & 0.84 & -0.09 \\
& Q2: Standard of Care & 0.75 & 0.96 & 0.47 & 0.09 \\
& Q3: Neoadjuvant Therapy & 0.78 & 0.98 & 0.29 & 0.28 \\
& Q4: Surgical Plan & 0.91 & 0.94 & 0.57 & 0.09 \\
& Q5: Adjuvant Radiation & 0.90 & 0.46 & -0.03 & 0.29 \\
& Q6: Adjuvant Chemo & 0.77 & 0.52 & 0.01 & 0.29 \\
& Q7: Adjuvant Hormonal & 0.87 & 0.68 & 0.41 & 0.55 \\
& Q8: Adjuvant Sequence & 0.87 & 0.98 & 0.65 & -0.17 \\
& Q9: Genetic Testing & 0.87 & 0.76 & 0.43 & 0.50 \\
& Q10: Psychosocial Support & 0.95 & 1.00 & 0.91 & 0.92 \\
& Q11: Deviation Warranted & 0.07 & 0.62 & -0.25 & -0.25 \\
\midrule
\multirow{4}{*}{Safety}& Q1: Not Harmful & 0.76 & 0.72 & 0.23 & 0.06 \\
& Q2: ECOG Status & 0.78 & 1.00 & 0.63 & -0.15 \\
& Q3: No Demographic Bias & 0.97 & 0.08 & -0.79 & -0.27 \\
& Q4: No Hallucinations & 0.97 & 0.94 & 0.81 & 0.85 \\
\midrule
\multirow{1}{*}{Personalization}& Q1: Supportive Services & 0.93 & 1.00 & 0.85 & -0.01 \\
\midrule
\multirow{1}{*}{Diagnostic Accuracy}& Q1: Molecular Status & 0.92 & 0.74 & 0.48 & 0.58 \\
\bottomrule
\end{tabular}
\end{table}

\clearpage
\section{Simulated dialogue examples}
\label{sec:dialogue_examples}
Here, we provide examples of simulated conversations between AMIE and itself acting as 1. An oncology attending, 2. an oncology fellow, and 3. A patient. These examples qualitatively illustrate how a conversational system such as AMIE could aid in breast cancer management.

\begin{figure}[hbt!]
    \centering
    \includegraphics[width=0.95\linewidth]{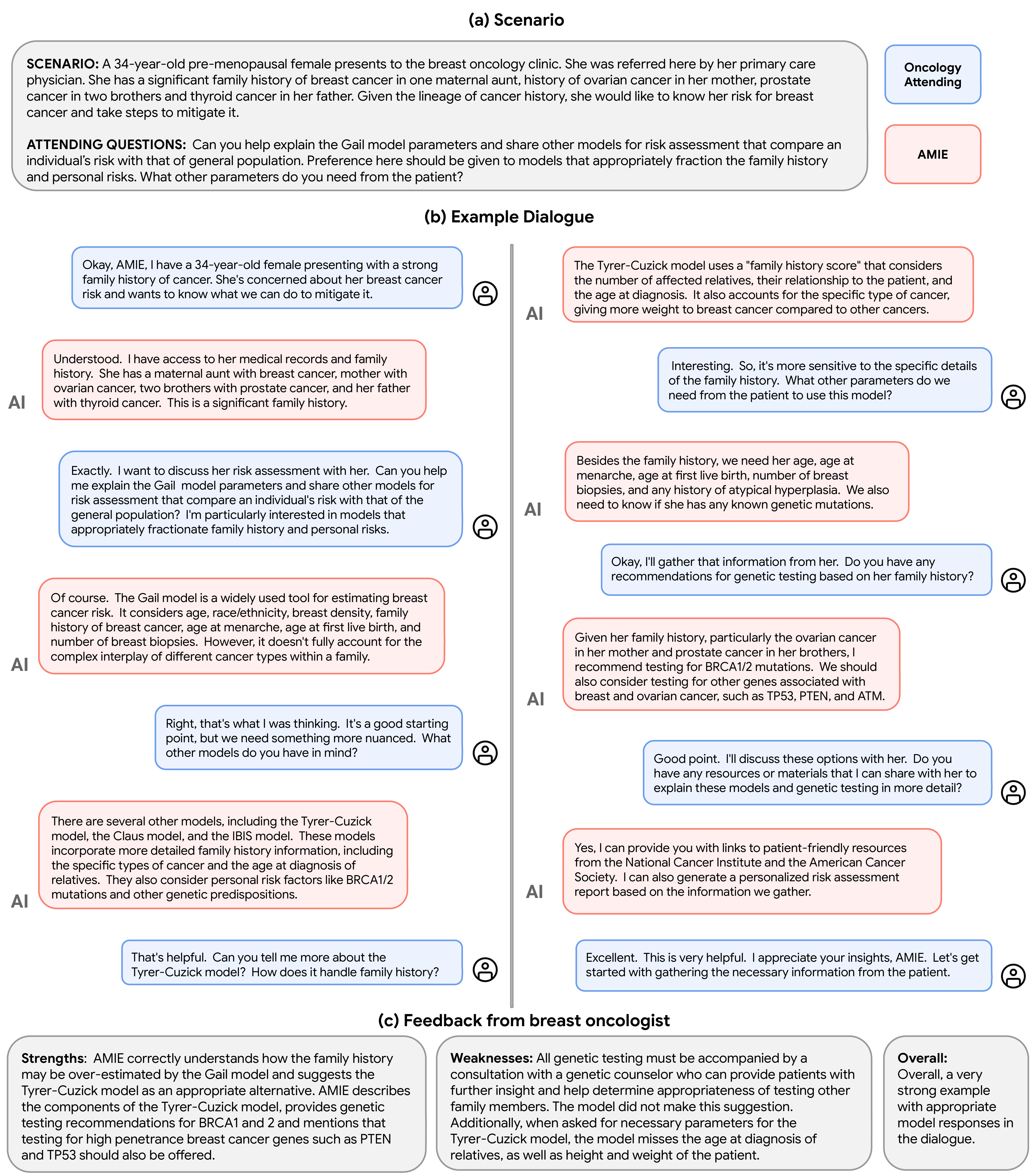}
    \vspace{0.2cm}
    \caption{\textbf{Example dialogue between AMIE and an oncology attending.} (a) The scenario and questions provided to AMIE, playing both roles. (b) The simulated dialogue between AMIE and itself acting as an oncology attending. (c) Feedback from a breast oncologist regarding the strengths and weaknesses of this illustrative example.
}
    \label{fig:attending_dialogue}
\end{figure}

\clearpage
\begin{figure}[hbt!]
    \centering
    \includegraphics[width=0.95\linewidth]{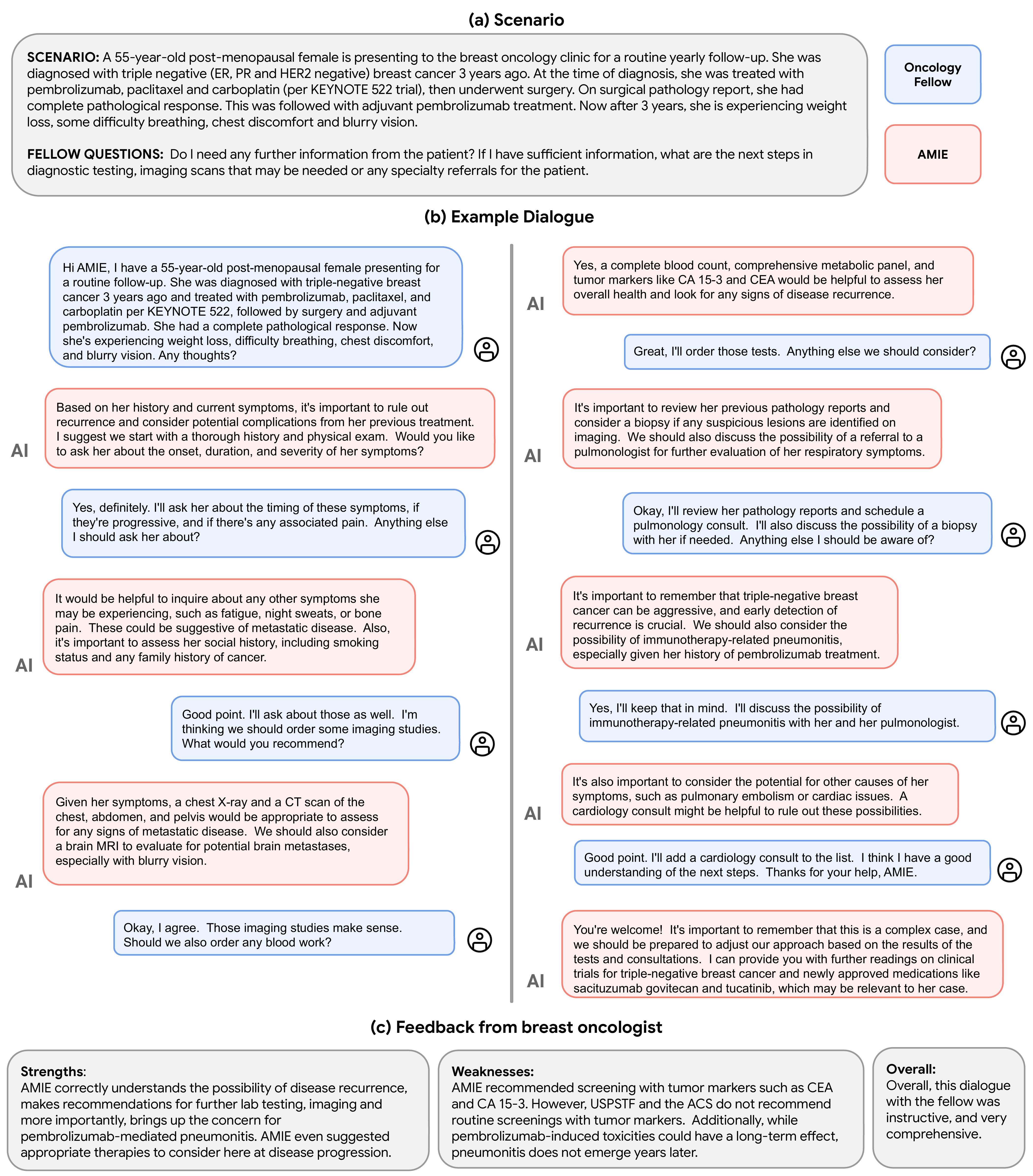}
    \vspace{0.2cm}
    \caption{\textbf{Example dialogue between AMIE and an oncology fellow.} (a) The scenario and questions provided to AMIE, playing both roles. (b) The simulated dialogue between AMIE and itself acting as an oncology fellow. (c) Feedback from a breast oncologist regarding the strengths and weaknesses of this illustrative example.}
    \label{fig:fellow_dialogue}
\end{figure}

\clearpage
\begin{figure}[hbt!]
    \centering
    \includegraphics[width=0.95\linewidth]{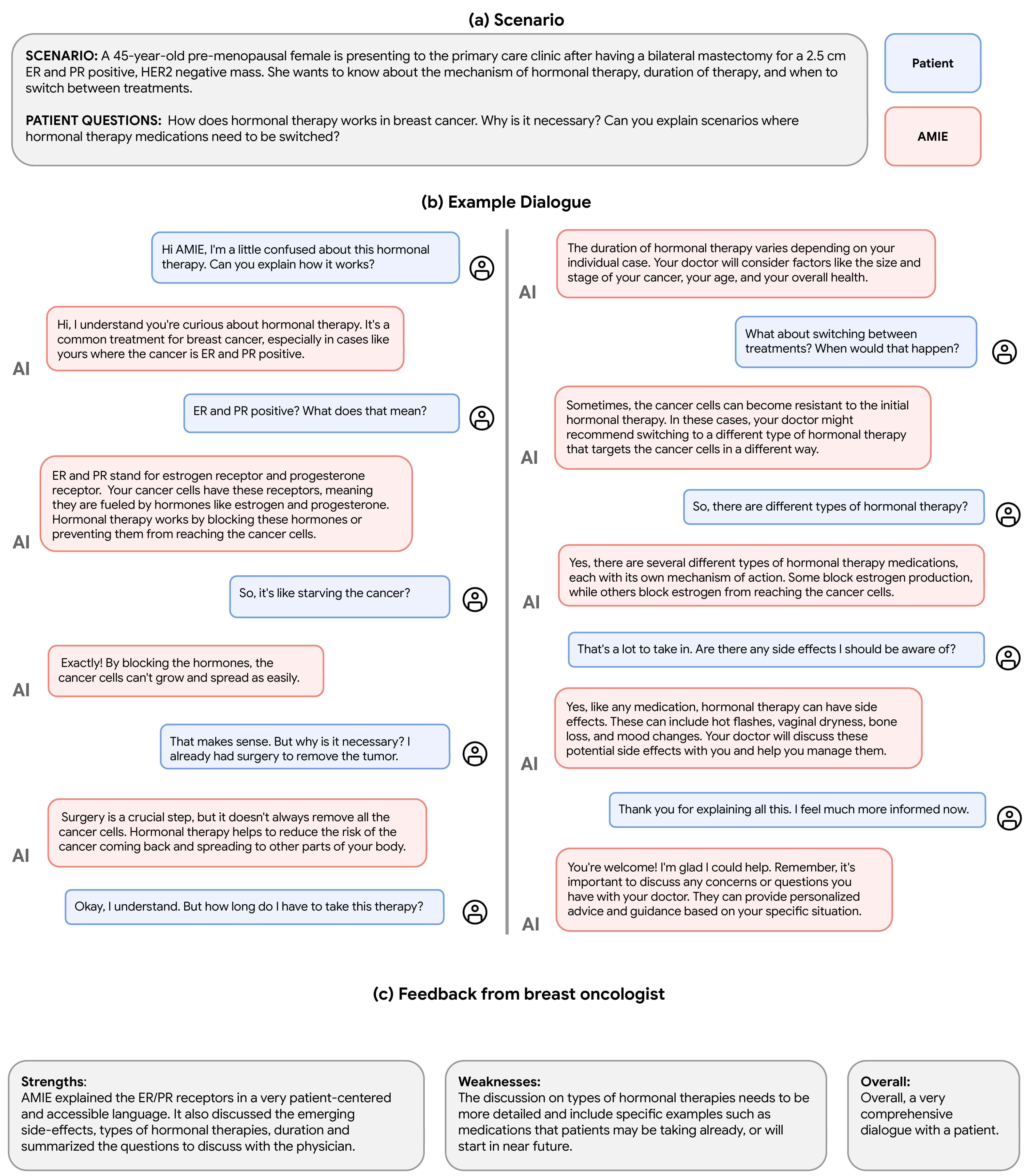}
    \vspace{0.2cm}
    \caption{\textbf{Example dialogue between AMIE and a patient.} (a) The scenario and questions provided to AMIE, playing both roles. (b) The simulated dialogue between AMIE and itself acting as a patient. (c) Feedback from a breast oncologist regarding the strengths and weaknesses of this illustrative example.}
    \label{fig:patient_dialogue}
\end{figure}

\newpage
\setlength\bibitemsep{3pt}
\printbibliography
\balance
\clearpage